\documentclass[12pt]{iopart}
\usepackage{bm}
\usepackage{epsfig}
\usepackage{amssymb}
\usepackage{iopams}
\begin{document}

\newcommand{\tw}{t_\mathrm{w}}
\newcommand{\rbf}{\boldsymbol{r}}
\newcommand{\st}{\mathcal{C}}

\newcommand{\len}{\ell}

\title[Fluctuation-dissipation relations in plaquette spin
systems]{Fluctuation-dissipation relations in plaquette spin systems
with multi-stage relaxation}

\author{Robert L Jack$^{1,2}$, Ludovic Berthier$^3$ and
Juan P Garrahan$^{4}$}
\address{$^1$ Rudolf Peierls Centre for
Theoretical Physics, University of Oxford, 1 Keble Road, Oxford, 
OX1~3NP, UK}
\address{$^2$ Department of Chemistry, University of California,
 Berkeley, California 94720-1460, USA}

\address{$^3$ Laboratoire des Collo\"{\i}des, Verres et Nanomat\'eriaux,
UMR 5587 Universit\'e Montpellier II \& CNRS, 34095 Montpellier
Cedex 5, France}

\address{$^4$ School of Physics and Astronomy, University of
Nottingham, Nottingham, NG7~2RD, UK}

\eads{\mailto{rljack@berkeley.edu}, 
\mailto{berthier@lcvn.univ-montp2.fr},
\mailto{juan.garrahan@nottingham.ac.uk} 
}

\begin{abstract} 
We study aging dynamics in two non-disordered spin models with multi-spin
interactions, following a sudden quench to low temperature.
The models are relevant to the physics of supercooled liquids. 
Their low temperature dynamics resemble those
of kinetically constrained models, and obey dynamical
scaling, controlled by
zero-temperature critical points. Dynamics in both models
are thermally activated, resulting in multi-stage
relaxation towards equilibrium. We study several two-time correlation
and response functions. We find that equilibrium
fluctuation-dissipation relations
are generically not satisfied during the aging regime, but deviations
from them are well described by fluctuation-dissipation ratios, as
found numerically in supercooled liquids.  These ratios are purely
dynamic objects, containing information about the nature
of relaxation in the models.  They are non-universal, and 
can even be negative as a result of activated dynamics. Thus,
effective temperatures are not well-defined in these models.
\end{abstract}

\maketitle

\section{Aging dynamics and plaquette models}

When a liquid is cooled through its glass transition,
its state changes: it becomes an amorphous solid \cite{ReviewsGT}.  
In this non-equilibrium state, physical
properties are not stationary, and the system displays aging
behaviour~\cite{Struik,Young}. A similar situation is encountered in many
different materials, ranging from disordered magnets to dense
granular media.  A full understanding of the non-equilibrium
glassy state remains a central theoretical challenge.  In this work,
we study the aging dynamics of two
spin models with multi-spin interactions \cite{Garrahan}. These 
finite-dimensional, non-disordered, dynamically heterogeneous 
systems have been of recent interest as models
for the glass transition in supercooled liquids 
\cite{BuhGar02,Jack05-spm,Jack-Garrahan}.

Theoretical studies of mean-field models have given
important insights into the aging dynamics of both structural
and spin glasses~\cite{CugKur}.
In these models, thermal equilibrium is never
reached, and aging proceeds by downhill motion in an increasingly flat
free energy landscape~\cite{laloux}.  
Time translational invariance is broken, and two-time correlation and
response functions depend on both their arguments.  The
fluctuation-dissipation theorem (FDT), which relates equilibrium
correlation and response functions, does not apply in the aging
regime. Instead, correlation and response are related by a
non-trivial fluctuation-dissipation ratio (FDR).  This led 
to the idea that aging systems might be characterised
by an effective temperature~\cite{CugKurPel97}, defined in terms of
the FDR.  Physically, relaxation in glassy
systems occurs in well-separated time sectors; it is easy to
imagine associating each sector with an effective 
temperature~\cite{kurchan}.
A thermodynamic interpretation of effective temperatures 
has also been suggested~\cite{CugKur,FraMezParPel98},  
related to the concept of replica symmetry breaking.
Taken together, these results lend considerable appeal to the
the mean-field description of aging 
(see \cite{CriRit03} for a review). 

However, there are many systems of physical interest in which
the dynamics are not of mean-field type, displaying both
activated processes and spatial
heterogeneity. Two examples are domain growth in disordered magnets
\cite{FisHus86}, and liquids quenched to below the glass transition 
\cite{ExpDH,ReviewsDH}. While some experiments
and simulations~\cite{CriRit03} seem to detect a
mean-field aging regime, theoretical studies have found ill-defined
FDRs~\cite{Cri-Vio}, non-monotonic response
functions~\cite{BuhGar02,nicodemi,kr,buhot,Garrahan-Newman}, observable
dependence~\cite{FieSol02,MayBerGarSol}, non-trivial FDRs without
thermodynamic transitions~\cite{BuhGar02}, and a subtle interplay
between growing dynamical correlation length scales and FDT
violations~\cite{barrat,kennett}; experiments have also detected
anomalously large FDT violations associated with intermittent
dynamics~\cite{exp}.  Moreover, at large times, aging often proceeds via
thermal activation, and it was recently shown \cite{Mayer06} that this
can lead to negative response and well-defined, but 
negative, FDRs.

In this work, our aim is to investigate 
further the effects of thermal activation
and dynamic heterogeneity on aging dynamics. To this end, we study two
different two-dimensional plaquette models with multi-spin 
interactions~\cite{Garrahan,BuhGar02,Jack05-spm,Jack-Garrahan,Garrahan-Newman,Lipowski,Newman-Moore}. On
one hand, they can be viewed as finite dimensional, 
non-disordered versions of $p$-spin models, 
and can be viewed as an 
attempt to transfer mean-field concepts to the finite dimensional 
world. 
On the other hand, they possess dual representations in terms of independent
plaquette excitations with constrained dynamics, and are therefore
directly related to the physics of kinetically constrained models
(KCMs) \cite{RitSol03}, with the advantage that both spin and
excitation degrees of freedom can be studied separately 
\cite{Other-plaquette}.  

In these models, low temperature relaxation
towards equilibrium proceeds via several distinct stages, each stage
being associated with a particular energy barrier.  While these
separate time scales are superficially reminiscent of the time
sectors found in mean-field spin glasses, their physical origin is 
quite different.  We will show that
the dynamics of plaquette models share some strong
similarities with mean-field aging dynamics, but also important
differences.  In addition, the dynamics of these models becomes
critical at low temperatures, where dynamical length scales
diverge. Our work therefore also pertains to the study of FDRs in
non-equilibrium critical dynamics \cite{CriRit03,CalGam}.

The first model that we study is the triangular plaquette model (TPM)
\cite{Garrahan,Garrahan-Newman,Newman-Moore}, defined for Ising spins, $s_i
=\pm1$, on the vertices of a triangular lattice. The Hamiltonian is
\begin{equation}
H_\mathrm{TPM} = -\frac{1}{2} \sum_{1,2,3 \,\, \in 
\,\, \nabla} s_1 s_2 s_3,
\label{htpm}
\end{equation}
where the sum is over the downward pointing triangular plaquettes of
the lattice.  We work with periodic boundary conditions in a system
of linear size $L$ of the form $L=2^k$. 
The dynamics involve single spin flips, with Glauber rates.

It is useful to define the binary dual plaquette variables by
\begin{equation}
n_i\equiv \frac{1}{2}(1-s_1 s_2 s_3),
\label{triangle}
\end{equation} 
where the three spins lie on the vertices of the $i$th
downward-pointing triangular plaquette.  
The thermodynamics of the system are those of the $L^2$
non-interacting dual spins $n_i$. Combining
(\ref{htpm}) and (\ref{triangle}) we have, 
at temperature $T = \beta^{-1}$, the equilibrium
defect density, $\langle n_i \rangle_\mathrm{eq} =
(1+e^{\beta})^{-1}$, and an obvious connection to kinetically
constrained models~\cite{Garrahan,Garrahan-Newman}.

\begin{figure}\hfill
\epsfig{file=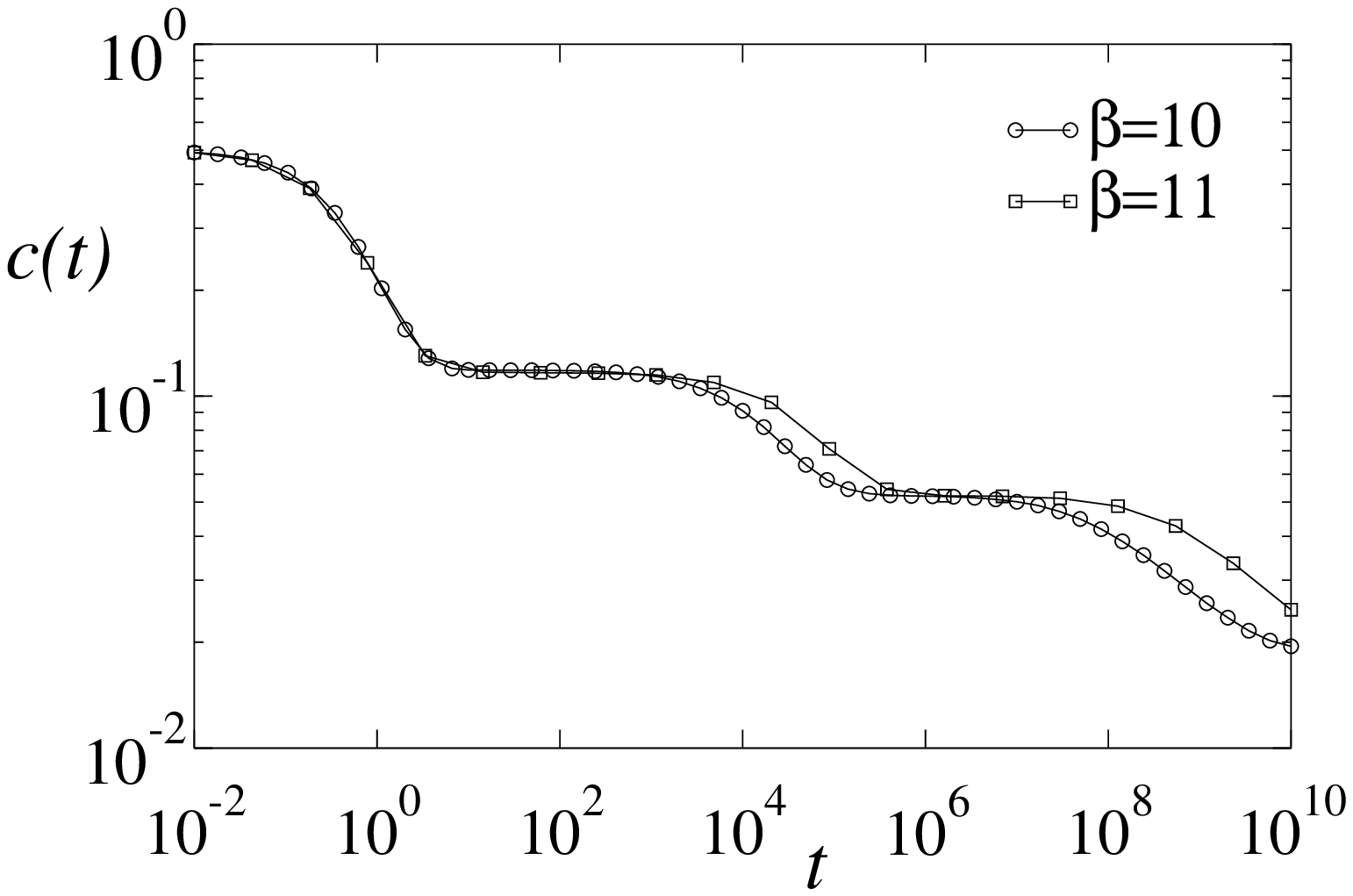,width=0.45\columnwidth}
\epsfig{file=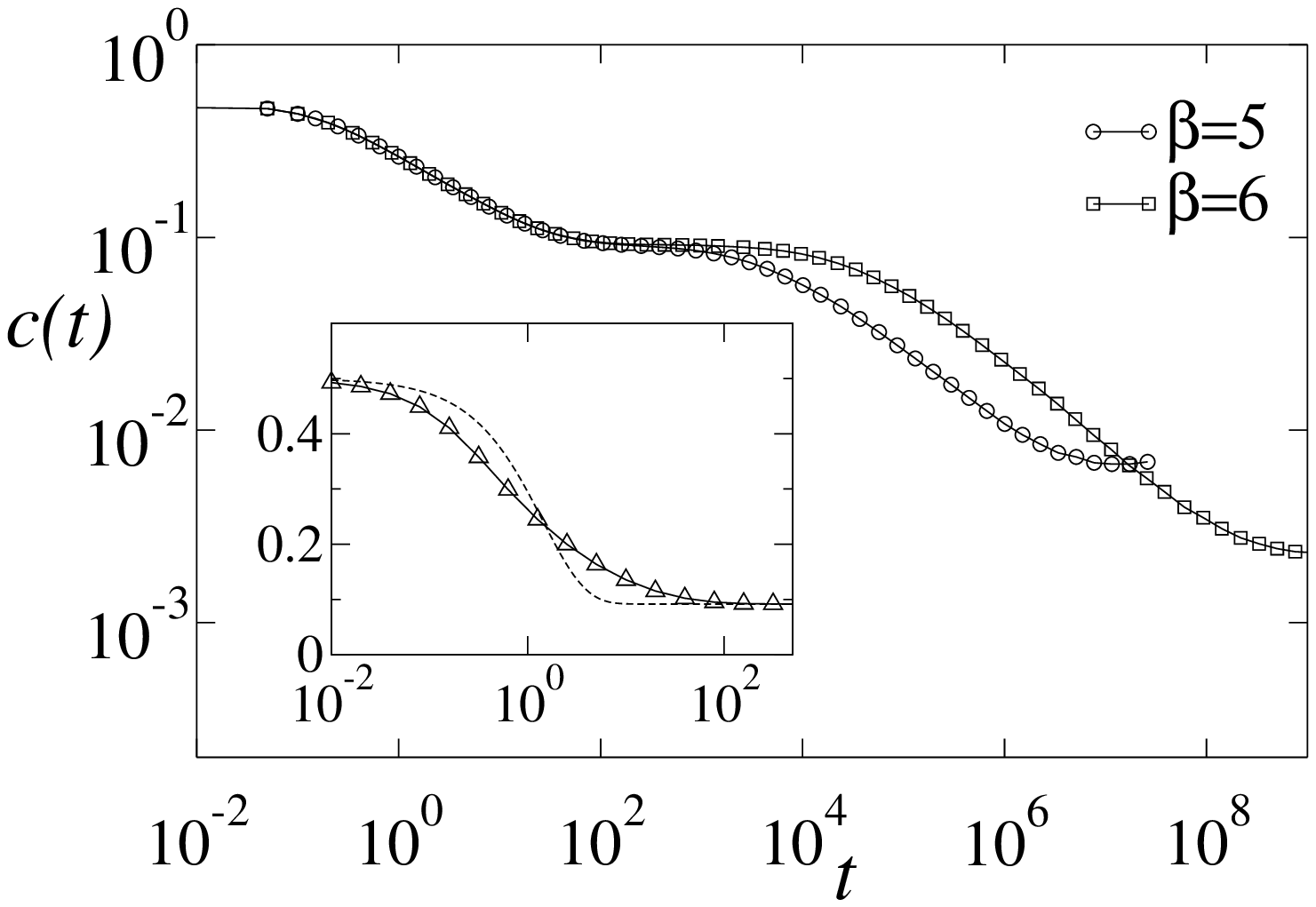,width=0.45\columnwidth}
\caption{Relaxation of the energy density in the TPM (left) and SPM
(right), after a quench from infinite temperature to $T=\beta^{-1}$. 
In the TPM, the relaxation proceeds in a hierarchical way, with stages
corresponding to integer energy barriers
\cite{Jack05-spm,Garrahan-Newman}. The system has not yet equilibrated:
this will occur when $c(t)=(1+e^\beta)^{-1}$.
In the SPM, the decay has just two stages; the second 
plateau in the decay of $c(t)$ represents equilibration
of the system.  The first stage of relaxation
has no energy barriers, and corresponds to zero
temperature dynamics (a quench to $T=0$ is shown in the inset, and
compared with a simple exponential as a dashed line);
the second step is characterised by
activated reaction-diffusion behaviour \cite{Jack05-spm}.}
\label{fig:tpm_spm}
\end{figure}

The TPM has a hierarchical free energy landscape: as the temperature
is lowered, the system falls into ever deeper potential wells, and the
equilibrium relaxation time diverges as $\tau \sim e^{\beta^2/\ln3}$
\cite{Jack-Garrahan}. This behaviour resembles that of
the one-dimensional East model~\cite{EastModel}, or its generalization 
in higher dimensions~\cite{nef1,nef2}.
The hierarchy of energy barriers in
the TPM is clear from the behaviour on a quench: in
Fig.~\ref{fig:tpm_spm}, we plot the defect (or energy) density
\begin{equation}
c(t)\equiv\langle n_i(t) \rangle,
\end{equation}
as a function of time, 
after a quench from infinite temperature (the average is over
initial conditions and over thermal histories).
The energy barriers relevant for the quench take integer values, so 
the decay of the energy becomes a function
of the scaling variable $T\log t$, as long as $t$ 
remains smaller than the equilibration time~\cite{SolEvans}.

Our second system is the square plaquette model (SPM)
\cite{Lipowski,Garrahan}, in which
we define Ising spins $s_i=\pm 1$ on
a square lattice, with
\begin{equation}
H = -\frac{1}{2}\sum_{1,2,3,4 \,\, \in \,\, \square} s_1 s_2 s_3 s_4,
\end{equation}
where the sum is over the elementary plaquettes of the square lattice.
For the SPM, we define the binary dual variables by
\begin{equation}
n_i \equiv \frac{1}{2} (1- s_1 s_2 s_3 s_4),
\end{equation}
where the four spins lie on the vertices of the $i$th square
plaquette.  We use the same symbol, $n_i$, for the dual spins in both 
plaquette models.  The thermodynamics are again those of
non-interacting dual spins, but constraints from the boundary
conditions cannot be avoided in this model \cite{Jack-Garrahan}. 
These are minimised by our
use of periodic boundary conditions, but it is necessary to use linear
system sizes $L$ significantly greater than the inverse density
of defects, $L \langle n_i \rangle \gg 1$. This is a rather strong
constraint at low temperature. In the thermodynamic limit, the
equilibrium defect density is the same as that of the TPM, $\langle
n_i \rangle_\mathrm{eq}=(1+e^{\beta})^{-1}$.

The low temperature behaviour in the SPM is that of dilute point defects. 
In equilibrium, they diffuse at a rate that is proportional to $e^{-3\beta}$;
similar activated diffusion occurs in the Fredrickson-Andersen (FA)
model \cite{FAModel}.  This process is mediated by pairs of defects
that diffuse rapidly along one-dimensional paths that are aligned with the
lattice axes.  Where the TPM has a hierarchical relaxation towards
equilibrium, the SPM has just two stages. The behaviour on a
quench is shown in Fig.~\ref{fig:tpm_spm}.  The second stage is a
reaction-diffusion process with activated diffusion, and the decay is a
function of the rescaled time $t e^{-2 \beta}$ \cite{BuhGar02,Jack05-spm}.  
The early time
regime of the SPM has no energy barriers. It corresponds to
relaxation into a jammed state at zero temperature. This relaxation is
frustrated by entropic effects leading to a non-exponential relaxation
of the energy density, as shown in the inset of
Fig.~\ref{fig:tpm_spm}.

Earlier work on aging dynamics in the TPM and SPM 
\cite{BuhGar02,Garrahan-Newman} suggested that the
combination of activated dynamics and hierarchical relaxation leads
to novel and intriguing behaviour for response functions and FDRs.
Here, we study these FDRs by computer simulations.  We investigate 
spin and defect observables, including
both local and spatially dependent correlations; we make use of a 
recently developed method for direct measurement of response functions 
in Monte Carlo simulations~\cite{Chatelain04,Ricci}. 

We find a rich
structure in the FDRs, with qualitatively different behaviour
in different stages of the relaxation. We observe 
FDRs close to unity for regimes in which
energy conserving processes allow the system to explore configuration 
space efficiently; correlations may also relax
by irreversible processes, whose rates are independent of the perturbing 
field; in this case the FDR is close to zero. The relative
rates for these processes reflect local properties of the 
free energy landscapes in these systems \cite{laloux}.  
We also observe
negative FDRs, coming from activated aging behaviour, 
as in \cite{Mayer06}.
 
The rest of the paper is
organized as follows: in Section II we study FDRs in the triangular
plaquette model; in Section III we do the same for the square
plaquette model; Section IV gives a brief summary 
of our results and their implications.  
Details of the no-field method for measuring
response functions are provided in \ref{app:chat}.

\section{Fluctuation-dissipation relations in the triangular plaquette model}

\subsection{Spin observables}

We begin by considering correlation and response functions for spins
in the TPM. The two-time spin autocorrelation function is
\begin{equation}
C_s(t,\tw) = \langle s_i(t) s_i(\tw) \rangle,
\label{equ:Cs}
\end{equation}
and its conjugate response is
\begin{equation}
\chi_s(t,\tw) = \frac{\mathrm{d}}{\mathrm{d}(\beta h_i)}
\langle s_i(t) \rangle,
\label{equ:chis}
\end{equation}
where $h_i$ is the strength of a magnetic field that acts on
site $i$ between times $\tw$ and $t$.
Averages are performed over
realisations of the initial condition and the thermal history, as above.  
For these observables, the FDR, $X_s(t,\tw)$, is defined by
\begin{equation}
\left.\frac{
\partial \chi_s(t,\tw)}{\partial 
\tw}\right|_t = -X_s(t,\tw) 
\left.\frac{
\partial C_s(t,\tw)}{\partial 
\tw}\right|_t 
.
\label{equ:def_X}
\end{equation}
In equilibrium, the fluctuation-dissipation theorem states that
$X(t,\tw)=1$ for all $t$ and $\tw$. However, away from
equilibrium, there are no such restrictions, as we shall see. 

Following Refs.~\cite{Chatelain04,Ricci}, two-time linear response functions
can be computed without using a perturbing field (see \ref{app:chat} 
for
details).  This allows direct access to the FDR.  
We emphasize that the derivatives in Eq.~(\ref{equ:def_X}) are
with respect to $\tw$.  For this reason, we present
data for fixed time $t$, as a function of $\tw$, unlike 
previous work on the same models
where the opposite convention was used~\cite{BuhGar02,Garrahan-Newman}.
We shall see that this difference qualitatively affects 
the results and their interpretation.

The dependence of the FDR on $\tw$ can be replaced by
a parametric dependence on the value of the correlation:
\begin{equation}
X(q,t) \equiv \left. X(t,\tw) \right|_{C(t,\tw)=q}. 
\label{equ:X_qt}
\end{equation}
That is, working at fixed time $t$, the waiting time $\tw$ is
parametrised by the value of $C(t,\tw)$.  A parametric plot of
$\chi(q,t)$ versus $q$ (an ``FD plot'') has local gradient $-X(q,t)$,
corresponding to the FDR defined in Eq.~(\ref{equ:def_X}). FD plots 
were suggested by the study of mean-field models where 
Eq.~(\ref{equ:X_qt}) simplifies to a single argument function 
$X(q,t)= {\cal X}(q)$. This asymptotic property has not been confirmed 
beyond mean-field~\cite{CriRit03}.

\begin{figure} \hfill
\epsfig{file=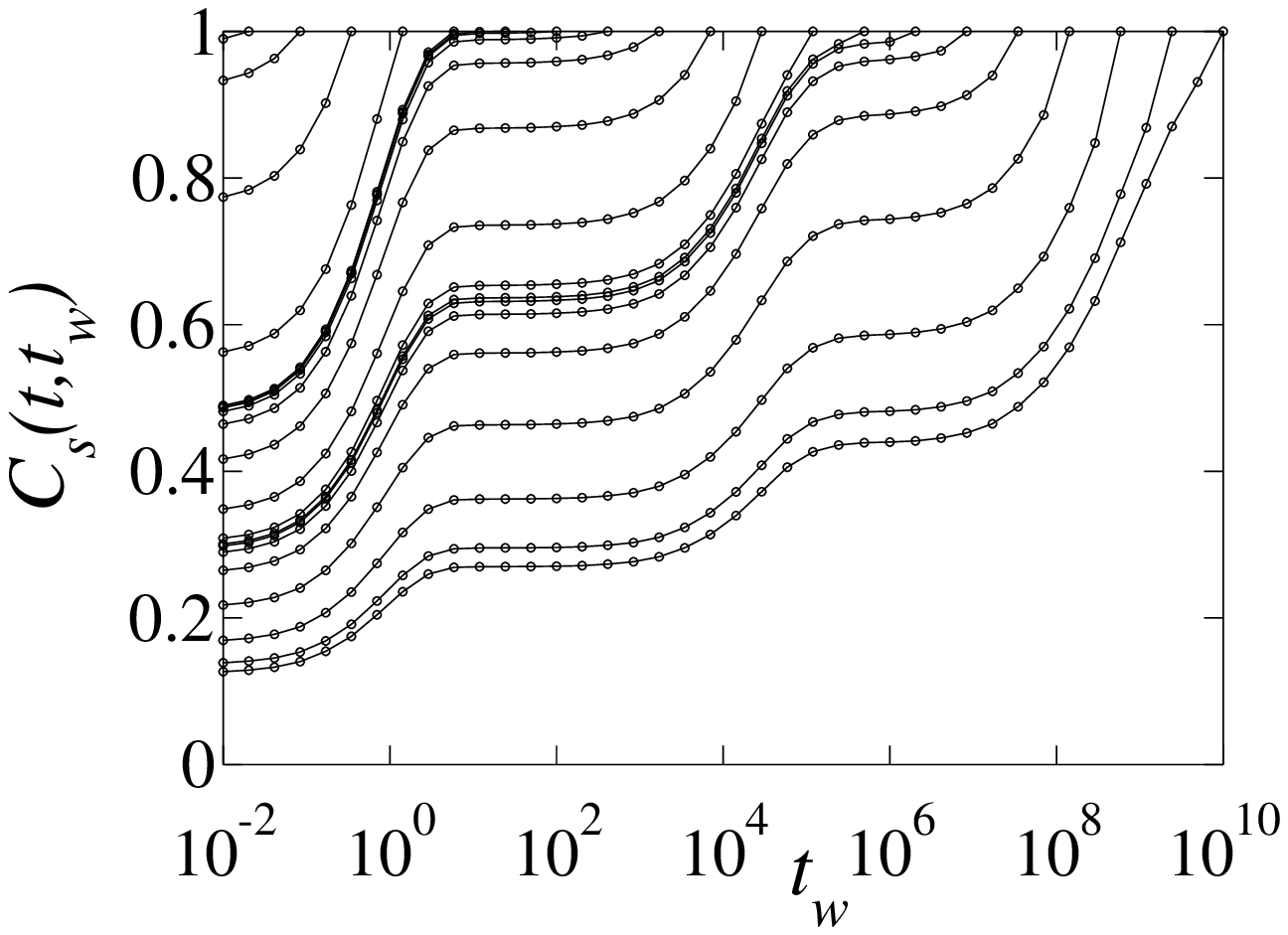,width=0.32\columnwidth}
\epsfig{file=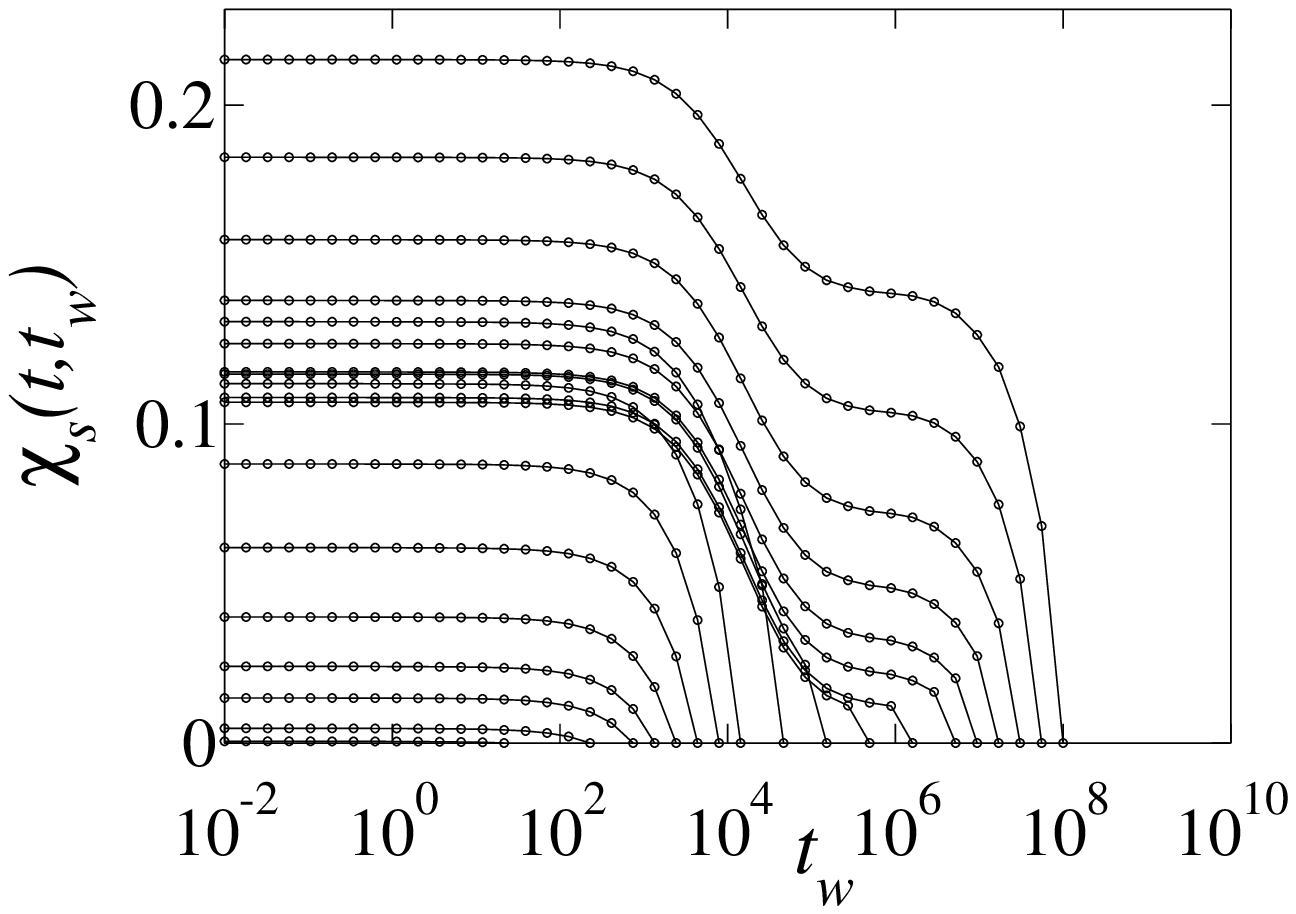,width=0.32\columnwidth}
\epsfig{file=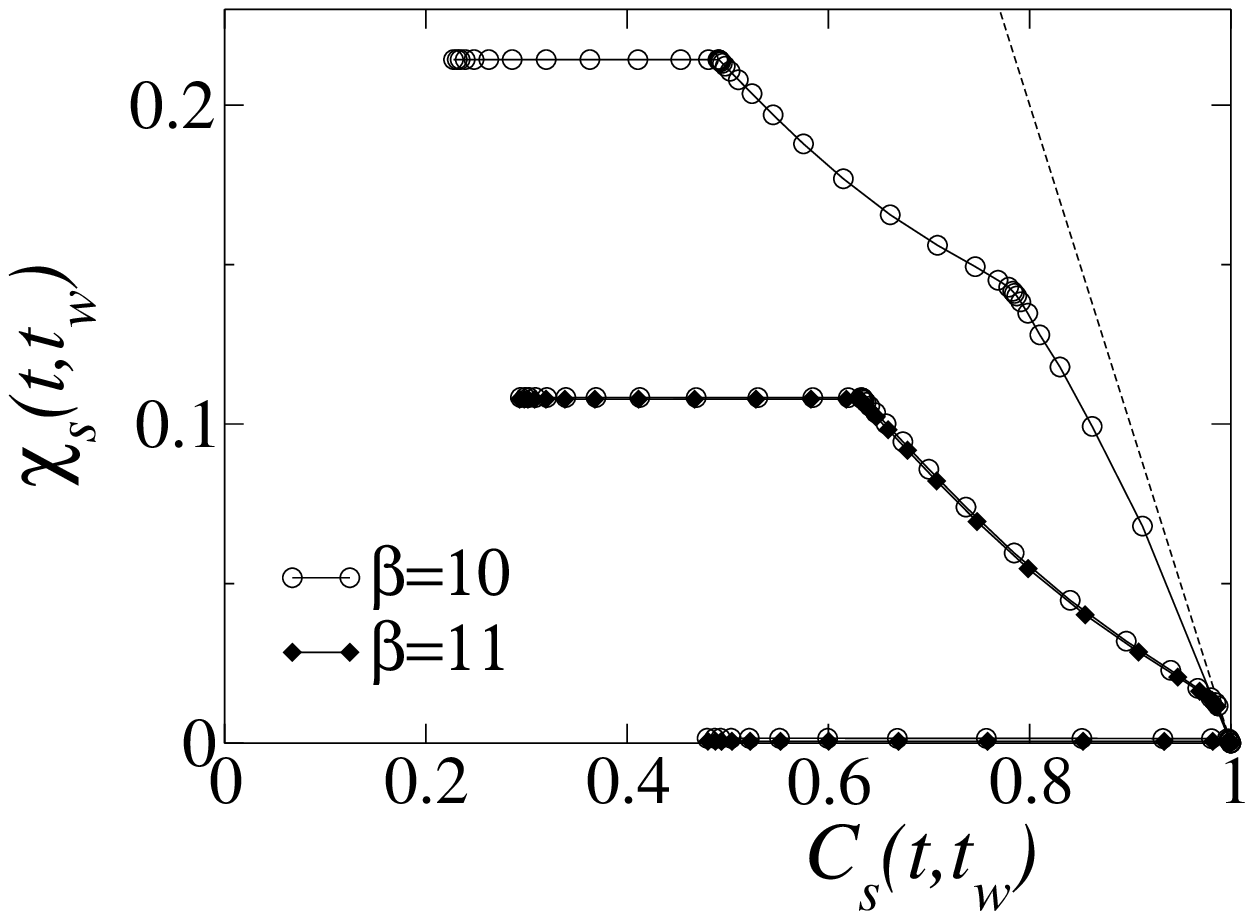,width=0.32\columnwidth}
\par\vspace{12pt}\hfill
\epsfig{file=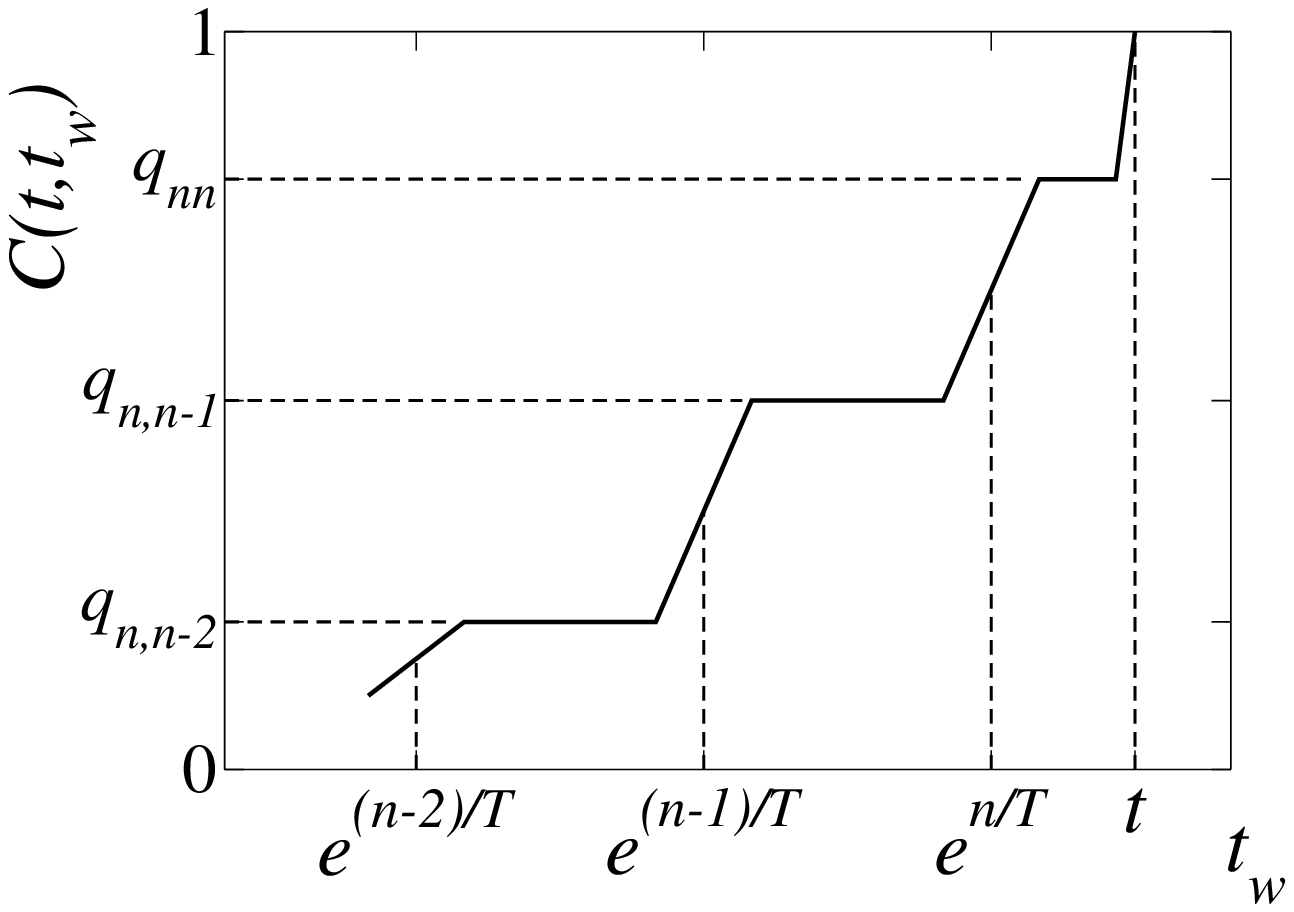,width=0.32\columnwidth}
\epsfig{file=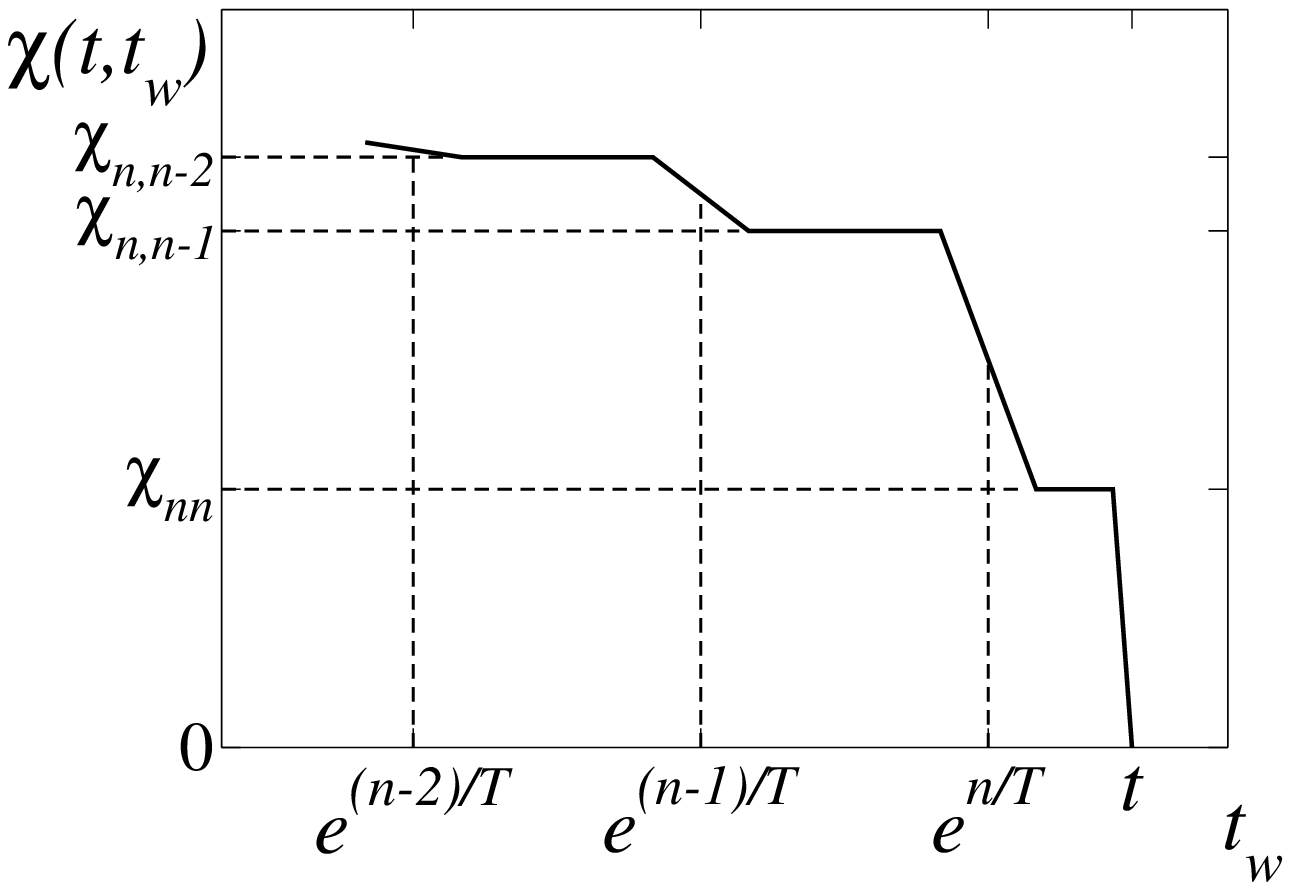,width=0.32\columnwidth}
\epsfig{file=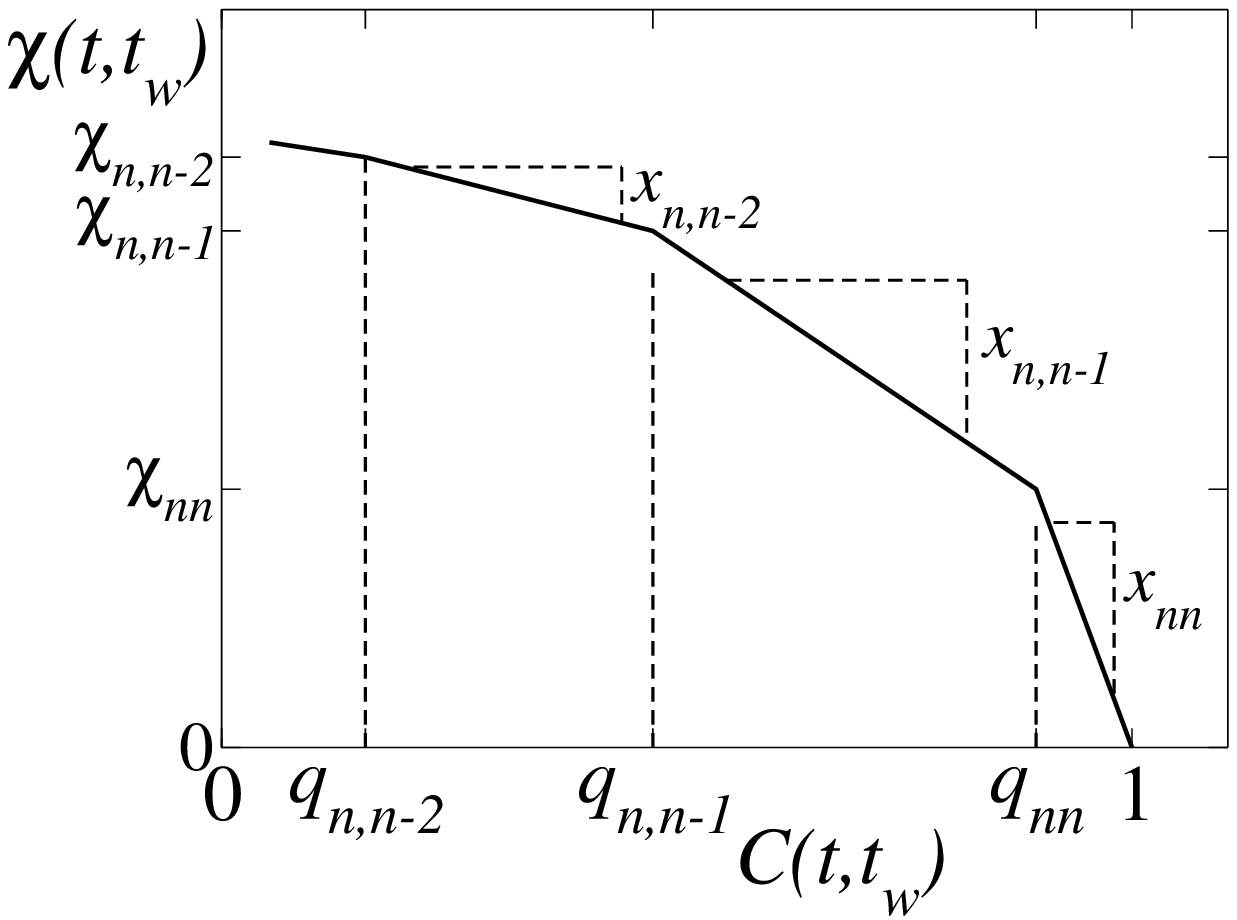,width=0.32\columnwidth}
\caption{(Top) Two-time spin autocorrelations (left), response functions 
(centre) and parametric FD plots (right)
in the TPM at $\beta=10$. The correlation and
response are plotted against $\tw$, for various $t$. The 
FD plot (top right) uses $\tw$ as a parametric variable; the three
traces at $\beta=10$ are $t=(70,1.6\times10^6,10^8)$. 
The equilibrium FD relation,
$\chi(t,\tw)=1-C(t,\tw)$, is shown as a dashed line. We also show
parametric plots for $\beta=11$ and 
$t=(70,1.7\times10^7)$.  Since these values of $t$ are within
plateaux of $c(t)$, the FDR depends very weakly on temperature,
although the correlation and response have quite different time 
dependence (not shown).  (Bottom) Sketches of these plots,
illustrating the various definitions given in the text. Note
that the parameters $x_{nm}$ label the gradient of the parametric plot.}
\label{fig:tpm_cs}
\end{figure}

The behaviour of the two-time spin autocorrelation function of
Eq.~(\ref{equ:Cs}) is shown in Fig.~\ref{fig:tpm_cs}.  If $t$
is held constant, and $\tw$ 
decreased, the correlation decreases, in
stages that mirror the decay of the energy after the quench (recall
Fig.~\ref{fig:tpm_spm}).  The generic structure of the correlation is 
sketched in Fig.~\ref{fig:tpm_cs}.  

The plateaux in the energy, $c(t)$, come from the system being
trapped in metastable states.  
The dynamics are effective in exploring the metastable state:
in the $n$th such state, events corresponding to energy
barriers smaller than $n$ are very fast, while those corresponding to
energy barriers greater than or equal to
$n$ are frozen, until such times that the system
escapes from the metastable state.  Thus, the autocorrelation
function quickly relaxes to a value $q_{nn}$, which measures the
self-overlap of this metastable state. As $\tw$ is decreased further,
we arrive at a new plateau in the the two-time function. This measures
the mutual overlap between the plateau state $n$ and the previous
plateau state $n-1$.  We denote this quantity by $q_{n,n-1}$; earlier
plateaux in the two-time function give the mutual overlaps $q_{nm}$
with $m<n$, see Fig.~\ref{fig:tpm_cs}.

The two-time local integrated response, $\chi_s(t,t_w)$, is also
plotted in Fig.~\ref{fig:tpm_cs}.  
The stages of the dynamics are still visible, although 
the response decreases with increasing $\tw$,
unlike the autocorrelation function. Note also that
curves for different $t$ cross one another at short $\tw$.  This means
that response functions plotted at fixed $\tw$ for increasing $t$ are
non-monotonic, which illustrates why the use of
$t$ as a parametric variable in FD plots produces quite unusual results
\cite{Cri-Vio,Garrahan-Newman}. We sketch the behaviour of the
response function in Fig.~\ref{fig:tpm_cs}, and define the plateau
values of  the response function by $\chi_{nm}$. By analogy with the 
mutual overlap, the mutual response
$\chi_{nm}$ is the value of the response function measured
in the stage $n$, given that the field was continuously applied from 
stage $m$ to stage $n$.

If the response is now plotted parametrically against the
correlation, then each stage of the dynamics appears as a separate
segment on the FD plot. It is clear from Fig.~\ref{fig:tpm_cs} that
$X(q,t)\neq 1$ in the TPM, reflecting the fact that the system is well
out of equilibrium.  More importantly, it is apparent that the
segments of the parametric plot in that figure are close to straight
lines, so that FD plots are reducible to a discrete set of numbers.  
We denote the slopes
observed in the FD plots by $x_{nm}$, as sketched in Fig.~\ref{fig:tpm_cs}. 
We define
\begin{equation}
x_{nn} = \frac{\chi_{nn}}{1-q_{nn}},
\end{equation}
and 
\begin{equation}
x_{nm} = \frac{\chi_{nm} - \chi_{n,m+1}}{q_{n,m+1}-q_{nm}},
\label{equ:nm}
\end{equation}
for $n>m$.
These equations are a discretized version
of Eq.~(\ref{equ:def_X}). Physically, each stage $n$
of the hierarchy corresponds to a given timescale, $t_n \sim e^{n/T}$,
and a well-defined length scale, $\ell_n$, giving rise to
well-defined two-time (or rather `two-stage') correlation and response
functions and FDRs.  The whole dynamical behaviour is therefore
encoded in a discrete set of real numbers, $\{ q_{nm}, \chi_{nm}, x_{nm}
\}$, which obey certain relations, as we shall see in 
section~\ref{sec:indep}.
The FD plots found for spin observables have
similar shapes to the ones reported in Ref.~\cite{Mayer06} 
for the East model, but differ significantly from earlier 
work~\cite{Garrahan-Newman}. 

The first segment of the response always has a gradient very close to
unity, $x_{nn}\simeq 1$, consistent with quasi-ergodic 
(or quasi-equilibrium) dynamics within
the metastable plateau state.  We also note that the last plateau,
which corresponds to correlation and response between stages $n$ and
0, has $x_{n0} \simeq 0$. This is to be expected, since the first
stage of the quench only involves unactivated relaxation events, and
these are accepted with a
probability close to unity. Thus, their rate depends very weakly on any
perturbing field, and the response associated with this
completely irreversible dynamics is very weak (more precisely, it is
$\mathcal{O}(e^{-\beta})$, which does indeed vanish at
small temperature). This is very similar to 
the zero-FDR found in coarsening ferromagnets~\cite{coar1,coar2}.

We have discussed the data in Fig.~\ref{fig:tpm_cs} at a single
temperature. However, the scaling behaviour of the TPM with
temperature is well-understood~\cite{Jack-Garrahan,Garrahan-Newman}.  We
find that the plateau values of the energy, the self-overlaps $q_{nm}$
and the ratios $x_{nm}$ depend only very weakly on temperature.  As
the temperature is reduced towards zero, the stages of the
dynamics become more clearly defined and timescales more separated.
If we plot data as a function of the rescaled times,
$(\nu,\nu_\mathrm{w})=(T\log t,T\log \tw)$, 
then the plateaux fall on top of one another,
and the parametric plots quickly become independent of
temperature (for a given value of $\nu$). 
This assertion is confirmed numerically in Fig.~\ref{fig:tpm_cs},
where we show that 
parametric plots for different temperatures, 
$\beta=10$ and $\beta=11$, but similar $\nu$, perfectly superimpose.  

Mean-field studies have suggested the possibility 
to define effective temperatures from FDRs through~\cite{CugKurPel97}
\begin{equation} 
T_{\rm eff}(q) = \frac{T}{{\cal X}(q)}.
\end{equation}
Our finding of piecewise linear FD plots in Fig.~\ref{fig:tpm_cs}
apparently offers a nice illustration of the intuitive idea that 
well-separated relaxation timescales could lead to quasi-thermalization
at a given effective temperature within each time sector, 
as is the case in mean-field spin glasses~\cite{CugKur}. 
A further property derived from  mean-field models is that the value of 
$T_{\rm eff}$ is shared by all physical observables, physically
implying that different degrees of freedom have also thermalized between
themselves. We now investigate this issue.

In the TPM, the spatial dependence of two-spin correlation functions is
trivial, since both two-spin correlations and responses are
purely local. The symmetries of the model imply \cite{Garrahan}
\begin{equation}
\langle s_i(t) s_j(\tw) \rangle = \delta_{ij} C_s(t,\tw),
\label{equ:tpm_twospin:c}
\end{equation}
and
\begin{equation}
\frac{\mathrm{d}}{\mathrm{d}h_i} \langle s_j \rangle 
= \delta_{ij} \chi_s(t,\tw), 
\label{equ:tpm_twospin}
\end{equation}
which trivially implies that all two-spin correlation functions are 
equivalent.

In order to study spatial correlations, it would in fact 
be necessary to consider 
four-point functions \cite{FranzDPG99}; the relevant correlations 
were discussed in \cite{Jack-Garrahan}, but we restrict ourselves to
two-point functions in this paper.  Interestingly, (\ref{equ:tpm_twospin})
also implies that 
the dynamics of the total magnetization are trivially related to those
of local observables.  This remark is potentially relevant for supercooled
liquids, where static two-point correlations are also believed to
be decoupled from the microscopic physics of cooperative motion.
 
\subsection{Defect observables}

\begin{figure}
\hspace{72pt}\epsfig{file=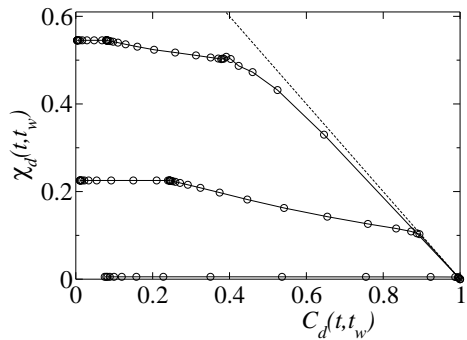,width=0.45\columnwidth}
\caption{Parametric FD plot for defects in the TPM, at $\beta=10$
and $t\in\{70,1.7\times10^6,10^8\}$, with $t$ decreasing
from top to bottom. The
structure is the same as for the spin observables
in Fig.~\ref{fig:tpm_cs},
but the FDRs are 
quantitatively different. 
The dashed line is the FDT: $\chi_d(t,\tw)=1-C_d(t,\tw)$.}
\label{fig:tpm_d}
\end{figure}

To understand the significance of the FDRs (and therefore effective 
temperatures) revealed by spin
observables, it is natural to ask whether different physical
quantities exhibit a similar behaviour. We therefore
consider the dynamics of the dual plaquette variables defined in
Eq.~(\ref{triangle}).  An excited plaquette, or ``defect'', is a
triangular plaquette for which $n_i=1$. Those defects become
increasingly dilute at low temperatures. We consider the defect
autocorrelation function,
\begin{equation}
C_d(t,\tw) = \frac{
\langle n_i(t) n_i(\tw) \rangle - \langle n_i(t) \rangle \langle
n_i(\tw) \rangle}{\langle n_i(t) \rangle [1-\langle n_i(t) \rangle]},
\end{equation}
and its conjugate response,
\begin{equation}
\chi_d(t,\tw) = \left[ 
\frac{\mathrm{d}}{\mathrm{d}(\beta g_{i})} \langle n_i(t)
\rangle \right]
\frac{1}{\langle n_i(t) \rangle [1-\langle n_i(t) \rangle]},
\end{equation}
where the perturbation $\delta H = -g_{i} n_i$ acts between $\tw$ and
$t$, and we evaluate the right hand side at $g_{i}=0$.  
We use connected correlation functions and 
we have normalised both correlation and response, so that 
$C_d(t,t)=1$, $C_d(t\to\infty,\tw)=0$, 
and the fluctuation-dissipation theorem for equilibrated systems reads
$\chi_d(t,\tw)=1-C_d(t,\tw)$.  

Overall, the results for defects are very similar to those for spins:
dynamic functions can be reduced to discretized overlaps and
responses. Moreover we find FD plots made up of linear segments, as shown
in Fig.~\ref{fig:tpm_d}.  However, we find that
the FDRs associated with the defects 
{\it differ quantitatively} from those of the spins, so we conclude 
that these numbers do not lead to useful effective temperatures.

These results are similar to those found
for the East model \cite{Mayer06}.  However, the low temperature
limits of the $q_{nm}$ and $x_{nm}$ can be estimated in the East
model by rather simple arguments~\cite{Mayer-prep}, 
whereas those in the TPM are
more complex, and contain information about the mechanisms of
relaxation in that system, as we now discuss.
In the East model, relaxation from 
stage $n-1$ to stage $n$ effectively consist of
selecting a random set of excitations and removing them. This
can be represented schematically by:\par
\begin{verbatim}
1 . . 1 . 1 . 1 . 1 . 1 . . 1 . .
1 . . 1 . . . 1 . . . 1 . . 1 . .
1 . . . . . . 1 . . . . . . 1 . .
\end{verbatim}
where we show configurations of the East model's defect variables, at three 
times, with the latest time at the bottom.  Neighbouring defects in the 
first state are separated by gaps of length at least 1, so it 
representative of the first
plateau. The second and third states are representative of the 2nd and
3rd plateaux respectively (gaps at least 2 and 4).
We can see that
\begin{equation}
\langle n_i(t) n_i(\tw) \rangle \simeq \langle n_i(t) \rangle, 
\end{equation}
so that
\begin{equation}
q_{nm}^\mathrm{(East)} 
\simeq  q_{nm}^{\rm max} \equiv
\frac{1-\langle n_i(\tw) \rangle}{1-\langle n_i(t) \rangle},
\label{equ:q_East}
\end{equation}
where $t$ is a time within the $n$th plateau of $c(t)$,
and $\tw$ a time within the $m$th plateau.

\begin{figure}
\hspace{72pt}\epsfig{file=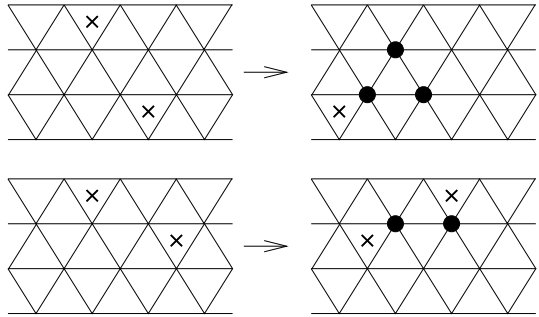,width=0.45\columnwidth}
\caption{Processes in the TPM requiring an energy barrier of 1
unit. Spins are defined on the vertices of the lattice. The $\times$ 
signs mark defects.  The black dots show spins 
that must flip for a transition between the left and right states. (Top)
Irreversible coagulation: a downward direction
in the energy landscape. (Bottom) Reversible (`diffusive') event:
the energy landscape is locally flat.}
\label{fig:tpm_dc}
\end{figure}

In the TPM, the situation is different, since there are both
reversible and irreversible moves taking place during the
relaxation. Irreversible coagulation events involve removing two
defects and adding a new one on a different site. 
Schematically, relaxation proceeds as:
\begin{verbatim}
1 . . 1 . . 1 . . 1 . 1 . . 1 .  
1 . . . 1 . . . . 1 . . . 1 . . 
\end{verbatim}
where pairs of defects combine to leave a single defect 
somewhere between the previous two. A more realistic 
representation of one such process is given in the top
panel of Fig.~\ref{fig:tpm_dc}.  Additionally, reversible moves
such as the one shown in the bottom panel of Fig.~\ref{fig:tpm_dc} can
occur.  In that particular example, two defects separated by distance
2 `rotate' around a given plaquette.
This process occurs on the same timescale as the coagulation event
shown in the same figure.  Both types of moves reduce
the self and mutual overlaps of the plateau states. Numerically we confirm
that  $q_{nm}^{\rm TPM} <
q_{nm}^\mathrm{(max)}$, but we are not able to estimate $q_{nm}^{\rm TPM}$
analytically.

The second difference between the TPM and East models is that 
dynamics at low temperatures in the East model
are those of irreversible events which allow no response,
so that defect FDRs vanish as $T \to 0$~\cite{Mayer-prep}.  
In the TPM, the $x_{nm}$
are finite numbers (except $x_{n0}\simeq 0$);
they become independent of temperature as
$T\to 0$.  
The rate for `coagulation' events is independent
of the local field at low temperatures, but `diffusive' events
do allow a response. For instance, if we take the
top left state of Fig.~\ref{fig:tpm_dc} as our initial condition, then
we end in the top right one regardless of the local field: the response
is zero.  On the other hand, 
if we start in the bottom left state then the final state
may be either the bottom left or bottom right one, with probabilities
weighted by the local fields: the response is finite. 
In the energy landscape picture \cite{laloux}, `diffusive' processes
are along directions in which the energy landscape is flat, and
`coagulative' processes correspond to downward paths in the energy
landscape.  We conclude that $x_{n,n-1}$ measures the
relative rates for `diffusive' and `coagulative' events during the
relaxation: a large value of $x_{n,n-1}$ is a 
sign that `diffusive' processes are dominating.  
We will confirm this
statement explicitly for the SPM in section \ref{sec:spm}, since the
relative rates for the two types of process can be controlled in
that model.

Finally, we note that the field $g_{i}$ is a local perturbation
to the temperature;
in the East model, this leads to negative FDRs, if the time $t$
is between plateaux of $c(t)$ \cite{Mayer06,Mayer-prep}. The
same behaviour occurs in the TPM, but we consider only 
the plateau FDRs, $x_{nm}$, in this article. These quantities
are positive (except for $x_{n0}\simeq0$).

\subsection{Independent steps approximation}
\label{sec:indep}

In the sketches of Fig.~\ref{fig:tpm_cs}, we reduced the 
behaviour of local dynamic observables to a discrete set of numbers.
We now show that if successive stages of the dynamics are
statistically uncorrelated, then several relations between these
numbers can be derived. These relations are satisfied by our
numerical data.

We use only the fact that probabilities for combinations of independent 
events factorise. We define $P_s(t,t')$ to be the probability that
a defect survives on a given site from time $t'$ to time $t>t'$.
If relaxation proceeds by independent stages then we expect
\begin{eqnarray}
P_s(t_n,t_m) &=&\prod_{n\geq n'>m} P_s(t_{n'},t_{n'-1})
,
\label{equ:inde}
\end{eqnarray}
where the times $\{t_i\}$ separate the stages.
We consider a situation where $t_n$ is a time within the $n$th plateau.

\begin{table}
\caption{Mutual overlaps and FDRs in the TPM at $\beta=10$. We
test the predictions of Eqs. (\ref{prod}) and (\ref{equ:x_indep}).
Consistent with those equations, 
we find $q_{31}\simeq q_{32} q_{21}$ and $x_{21}\simeq x_{31}$.
We estimate the uncertainty for the spin overlaps to be less than $10\%$
and those for the defect overlaps to be less than $20\%$; those
for the FDRs are given.  The $(*)$ indicates
that the defect value of 
$x_{31}$ was evaluated at $t=10^8$ (as in figure~\ref{fig:tpm_d});
the third stage of
relaxation was not complete at this time, but the FDR still seems
consistent with (\ref{equ:x_indep}).}
\label{tab:factor}
\begin{indented}
\item[] \begin{tabular}{@{}l|cccc|cc}
\br
& $q_{21}$ & $q_{32}$ & $q_{31}$ & ($q_{32}q_{21}$) 
& $x_{21}$ & $x_{31}$ \\ 
\mr 
Spins & $0.63$ & $0.43$ & $0.26\phantom{0}$ & 
(0.27)\phantom{1} & 
        $0.25\pm0.02$ & $0.21\pm0.04\phantom{(*)}$  
        \\
Defects & $0.25$ & $0.053$ & $0.009$ & (0.013) & $0.18 \pm 0.02$ & 
$0.17\pm0.03 (*)$ \\ \br
\end{tabular} \end{indented} \end{table}

A naive approximation for the survival probability 
would be to assume that $P_s(t,\tw)
\simeq \langle n_{it} n_{i\tw} \rangle / \langle n_{i\tw} \rangle$,
which is the probability that there is a defect on site $i$ at time
$t$, given that there was one on site $i$ at time $\tw$.  
In the East model, this approximation is appropriate: we know
that $q_{nm}=q_{nm}^\mathrm{max}$, and it follows that
$\langle n_{it_n}
n_{it_m} \rangle \simeq \langle n_{it_m} \rangle \prod_{n'=m+1}^n \langle
n_{it_{n'}} n_{it_{n'+1}} \rangle / \langle n_{it_{n'}} \rangle$.
On the other hand, the approximation fails if successive plateau
states are completely uncorrelated, in which case 
$\langle n_{it} n_{i\tw} \rangle
=\langle n_{it} \rangle\langle n_{i\tw} \rangle$. In this latter case, 
the
naive approximation fails because it includes in the survival probability 
events in which the defect on site $i$ is destroyed, and then replaced by 
a new defect. To take this into account, we estimate
instead
\begin{equation}
P_s(t,\tw) \simeq \frac{ \langle n_{it} n_{i\tw} \rangle - \langle
n_{it}\rangle \langle n_{i\tw} \rangle } {\langle n_{i\tw} \rangle (
1- \langle n_{i\tw} \rangle)}.
\label{equ:ps_factor}
\end{equation}
With this approximation, the factorisation of probabilities for
independent stages leads to
\begin{equation}
C_d(t,\tw) \simeq C_d(t,t_{n-1}) C_d(t_{n-1},\tw),
\label{equ:c_factor}
\end{equation}
which follows from (\ref{equ:inde}) and (\ref{equ:ps_factor}),
if the dynamics between times $t_{n-1}$ and $t$ are
statistically uncorrelated with the dynamics between times
$\tw$ and $t_{n-1}$. Times $t$ and $\tw$ are chosen 
within the $n$th and $m$th stages, respectively.  Correlators
for spin observables factorise in a similar way.
In terms of overlaps between plateau states, 
it follows that 
\begin{equation}
q_{nm} = \prod_{n\geq n'>m} q_{n',n'-1} 
.
\label{prod}
\end{equation}
This relation is satisfied both for the East model and for
completely uncorrelated stages.  The dynamics of the TPM are rather
slow, so that our data is limited, but we do obtain numerically that
$q_{31} \simeq q_{32} q_{21}$ for both the spin and defect
autocorrelations, see Table~\ref{tab:factor}. 

To establish the effect of the independent stage approximation
on the FDR, we consider the impulse response.
In the presence of an instantaneous field $h$ acting at time $\tw$, 
we expect the response to be given by the density of field-induced
defects at stage $n-1$, multiplied by the probability of these
defects surviving to time $t$. That is,
\begin{equation}
\frac{\mathrm{d}\langle n_i(t) \rangle_{h}}{\mathrm{d}h} = 
P_s(t,t_{n-1}) \frac{\mathrm{d}\langle n_i(t_{n-1}) \rangle_{h}}{\mathrm{d}h},
\end{equation}
which holds if
relaxation between $t_{n-1}$ and $t$ is independent of
that between $\tw$ and $t_{n-1}$, as before.
The derivative of $\chi(t,\tw)$ with respect to $\tw$ is
a response to an instantaneous field at $\tw$: taking
care of normalisation, we find that
\begin{equation}
\left.\frac{\partial\chi_d(t,\tw)}{\partial\tw}\right|_t 
= C_d(t,t_{n-1}) 
\left.\frac{\partial\chi_d(t_{n-1},\tw)}{\partial\tw}\right|_{t_{n-1}}.
\end{equation}
Again, the analysis for spin observables is similar.
Using the definition of the FDR, and Eq.~(\ref{equ:c_factor}),
we arrive at $
X(t,\tw)=X(t_{n-1},\tw),
$ and hence
\begin{equation}
x_{nm}=x_{n-1,m},
\label{equ:x_indep}
\end{equation}
for $n-1>m$. Thus, $x_{nm}$ is a function of $m$ only.
The `self FDR', $x_{mm}$, is close to unity. Physically,
the FDR only depends on the stage of the dynamics during which the 
instantaneous field 
is applied.  Our numerical results are 
again consistent with this analysis, see Table~\ref{tab:factor}.

\subsection{Spatial structure in defect FDRs}

In contrast to the spin degrees of freedom, defects have non-trivial
spatial correlations in the TPM (except at equilibrium, where defect 
correlations vanish at equal times).  
Defining the Fourier transform of the plaquette
field $n_{\bm{k}}(t) = N^{-1/2} \sum_i n_i(t) e^{i\bm{k}\cdot
\bm{r}_i}$, where $N$ is the number of plaquettes, 
we consider the correlator
\begin{eqnarray}
\tilde{C}_d(k,t,\tw) & = & \frac{\sum_{|\bm{k}'|\leq k} 
\langle n_{\bm{k}'}(t) n_{-\bm{k}'}(\tw) \rangle 
- \delta_{\bm{k}'} \langle n_{\bm{k}'} \rangle^2}{\sum_{|\bm{k}'|\leq k} (1)}
\nonumber\\
& = & \frac{1}{N} \sum_{ij} F_k(|\bm{r}_i-\bm{r}_j|) \left[
\langle n_i(t) n_j(\tw) \rangle - 
\langle n_i(t) \rangle\langle n_j(\tw) \rangle \right],
\label{correl}
\end{eqnarray}
where $F_k(|\bm{r}|)=\frac{\sum_{|\bm{k}'|\leq k} e^{i\bm{k}'\cdot\bm{r}}}
{\sum_{|\bm{k}'|\leq k} (1)}$ decays on a length scale of the order of
$k^{-1}$ from a value of unity at the origin. (Sums over
wave vector $\bm{k}$ are over the first Brillouin zone, with 
additional restrictions as specified.) Compared to the
apparently simpler correlator $\langle n_{\bm{k}}(t) n_{-\bm{k}}(\tw)
\rangle$, $\tilde{C}_d(k,t,\tw)$ attaches more weight to short distances;
measurements of this quantity
are therefore less affected by noise in the correlations at large
distances.  Moreover, $\tilde{C}_d(k,t,\tw)$ 
interpolates smoothly between the local
autocorrelation function when $k$ is large, and the global correlation
function of energy fluctuations at $k=0$.
Normalising as in the previous sections, we define
\begin{equation}
C_d(k,t,\tw) = \frac{\tilde{C}_d(k,t,\tw)}{\tilde{C}_d(k,t,t)}
\end{equation}
\begin{figure}\hfill
\epsfig{file=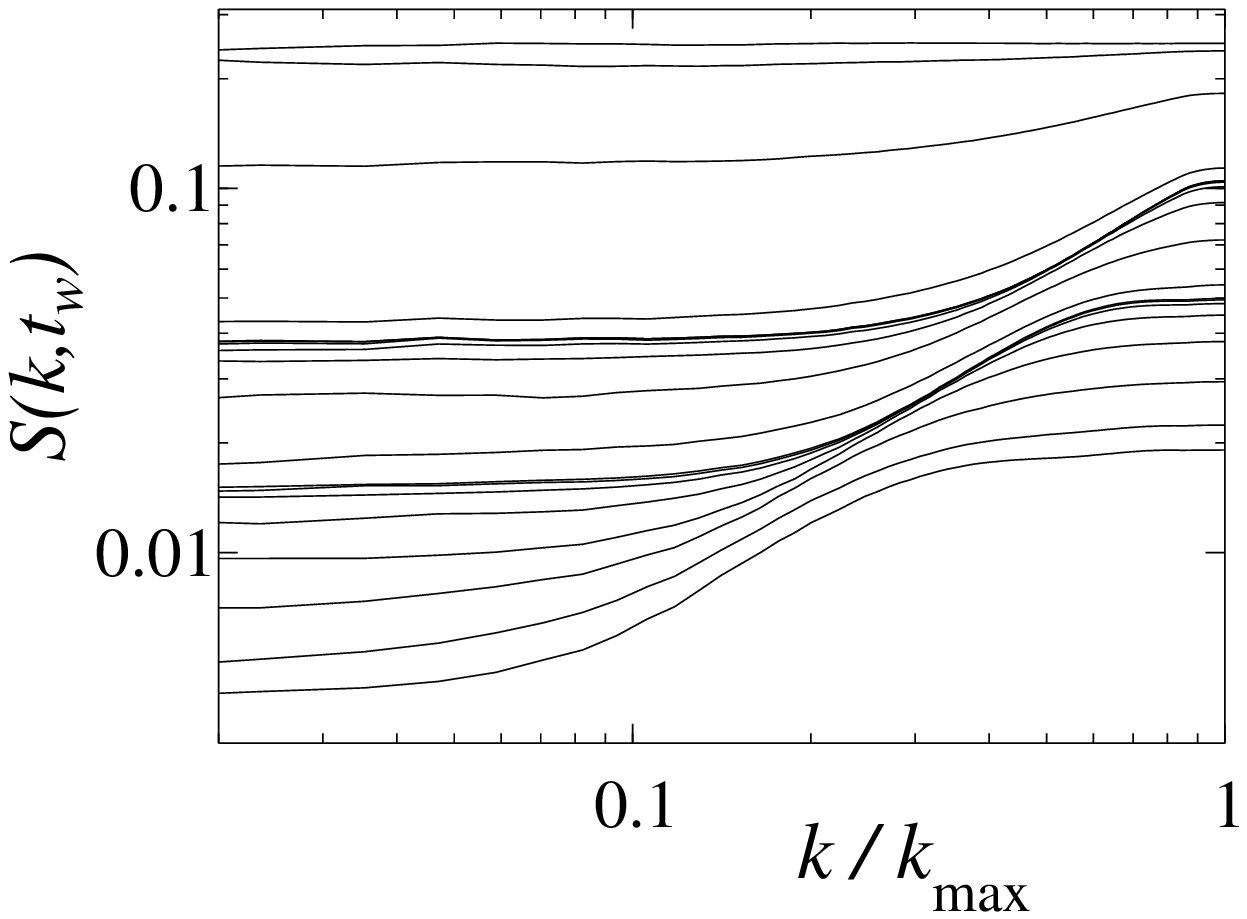,width=0.32\columnwidth}
\epsfig{file=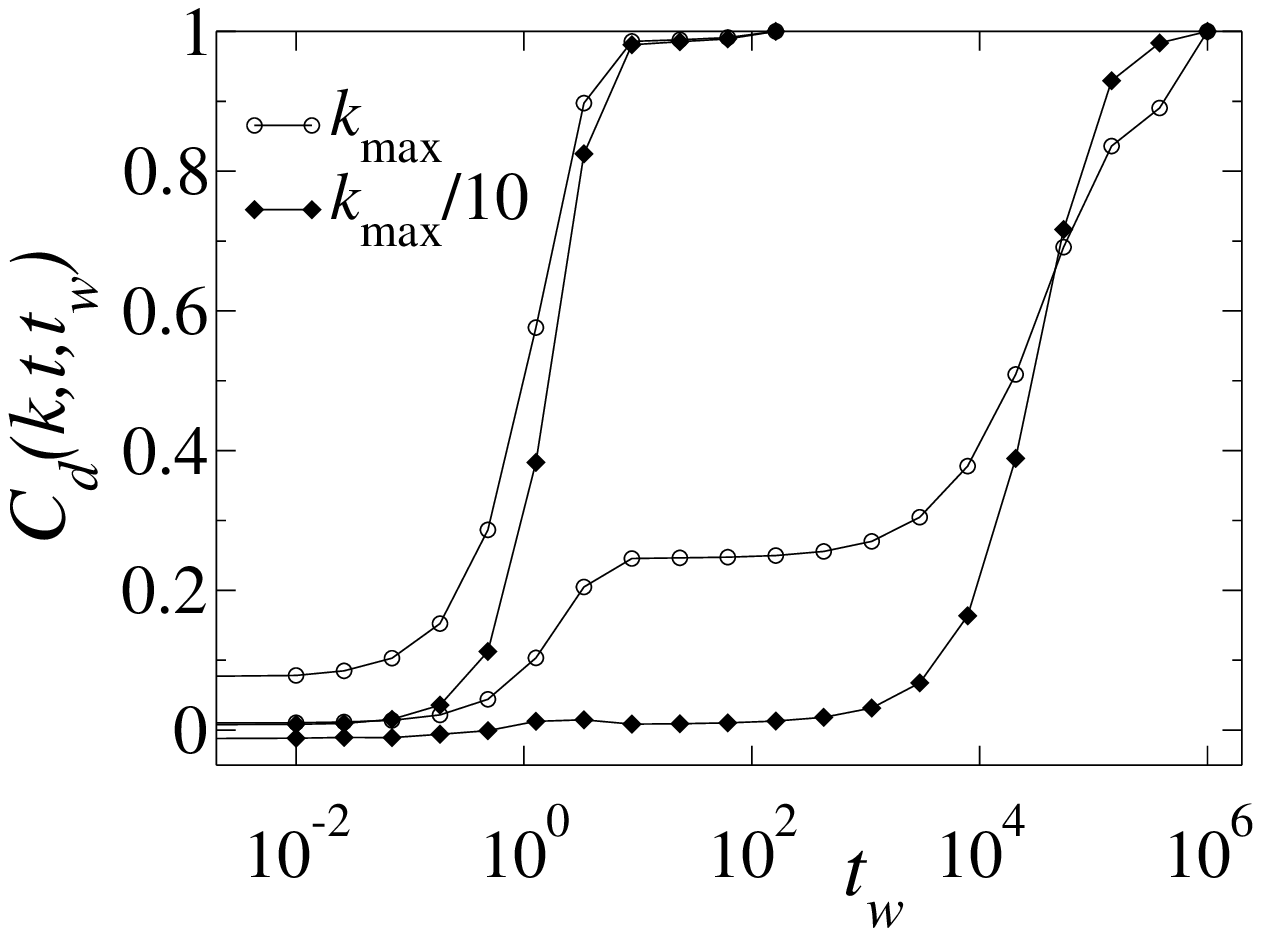,width=0.32\columnwidth}
\epsfig{file=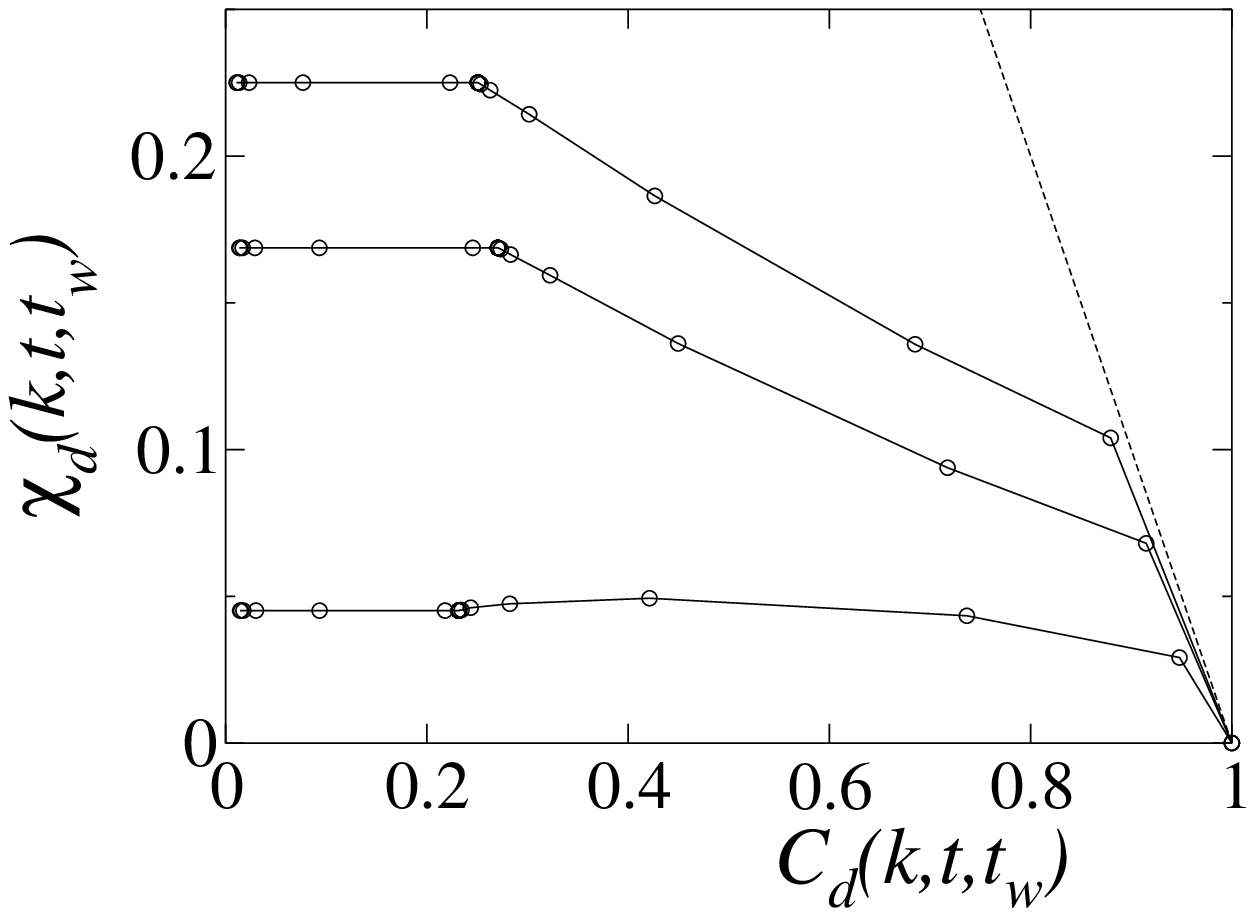,width=0.32\columnwidth}
\caption{
(Left) Structure factor of the defects, $S(k,\tw)$, at
$\beta=10$, and with $\tw$
logarithmically spaced between $10^{-2}$ and $10^6$ (from top to
bottom). A growing length scale is apparent, whose origin is the
increasing range of coagulative events at large $\tw$.
(Center) We show the correlation function of Eq.~(\ref{correl}), 
for $k\in\{k_{\rm max}, 
k_\mathrm{max}/10\}$, where $k_\mathrm{max}$ is the largest
wave vector in the Brillouin zone. 
We have lowered the temperature to $\beta=20$, and restrict ourselves
to the first and second stages of relaxation.
(Right) FD plots at $\beta=20$ for $k\in
\{k_{\rm max}$, $2k_{\rm max}/3$, $k_{\rm max}/2\}$ 
(from top to bottom) showing decreasing
response with decreasing $k$. The time is $t=2\times10^{10}$; at
this temperature then this time is 
within the second plateau of $c(t)$.}
\label{fig:tpm_cq}
\end{figure}

We begin with the case $t=\tw$, which defines the static structure
factor for the defects at time $\tw$:
\begin{equation}
S(k,\tw)\equiv \tilde{C}_d(k,\tw,\tw).
\end{equation}
We plot this function in Fig.~\ref{fig:tpm_cq}.  We see that
$S(k,\tw)$ is an increasing function of $k$, which means that defects
are surrounded by regions of reduced defect density. This is
consistent with a local coagulative process which induces 
some effective repulsion between defects.
There is a length scale
associated with the $k$-dependence of $S(k,\tw)$:  the time scale 
associated with coagulation of defects increases with their separation, 
so this length scale increases with $\tw$.

Moving to the time dependence of the correlation functions, we compare
correlators  at different wave vectors in the middle panel of
Fig.~\ref{fig:tpm_cq}. The plateau structure that is seen
in the local correlations is not present at small $k$. Instead, the
self overlaps of the plateau states approach unity, and the mutual
overlaps approach zero. This is to be expected: within the plateau
state, relaxation happens on small length scales, so
fluctuations on large length scales do not relax, and
self overlaps are large, for small $k$.  On the other hand, successive 
plateau states do not retain any memory of the average density in the
preceding plateau, so mutual overlaps at small $k$ are very small.

The normalised response conjugate to $C_d(k,t,\tw)$ is
\begin{equation}
\chi_d(k,t,\tw) = \frac{1}{C_d(k,t,t)}\frac{1}{N} 
\sum_{ij} F_k(|\bm{r}_i-\bm{r}_j|)
\frac{\mathrm{d}}{\mathrm{d}(\beta g_{j})}\langle n_i(t) \rangle
. 
\end{equation}
It measures the response to a random field which is correlated over a
length scale of $k^{-1}$.  We present FD plots for three wave vectors
in Fig.~\ref{fig:tpm_cq}, at a single time $t$ that is
within the second plateau of $c(t)$.  We find that the amplitude of 
the response functions quickly decrease when $k$ decreases and 
vanish as $k \to 0$. This can be simply inferred from
Fig.~\ref{fig:tpm_spm} since the energy density of each plateau states
is temperature independent.  Hence, the linear response of
the energy to a temperature change vanishes in the small temperature
limit. The situation is therefore similar to the one found in Ising 
or FA models~\cite{MayBerGarSol,Mayer06}: 
$k$-dependent FD plots smoothly interpolate
from the one obtained from local autocorrelation functions 
to the one obtained for global quantities at $k=0$, the crossover taking place
when $k \xi(\tw) \approx 1$, where $\xi(\tw)$ is a typical
length scale characterizing the correlations between defects in the system.

\section{Fluctuation-dissipation relations in the square plaquette model}
\label{sec:spm}

\subsection{Early time regime: zero temperature dynamics}
\label{sec:spm_init}

\begin{figure}\hfill
\epsfig{file=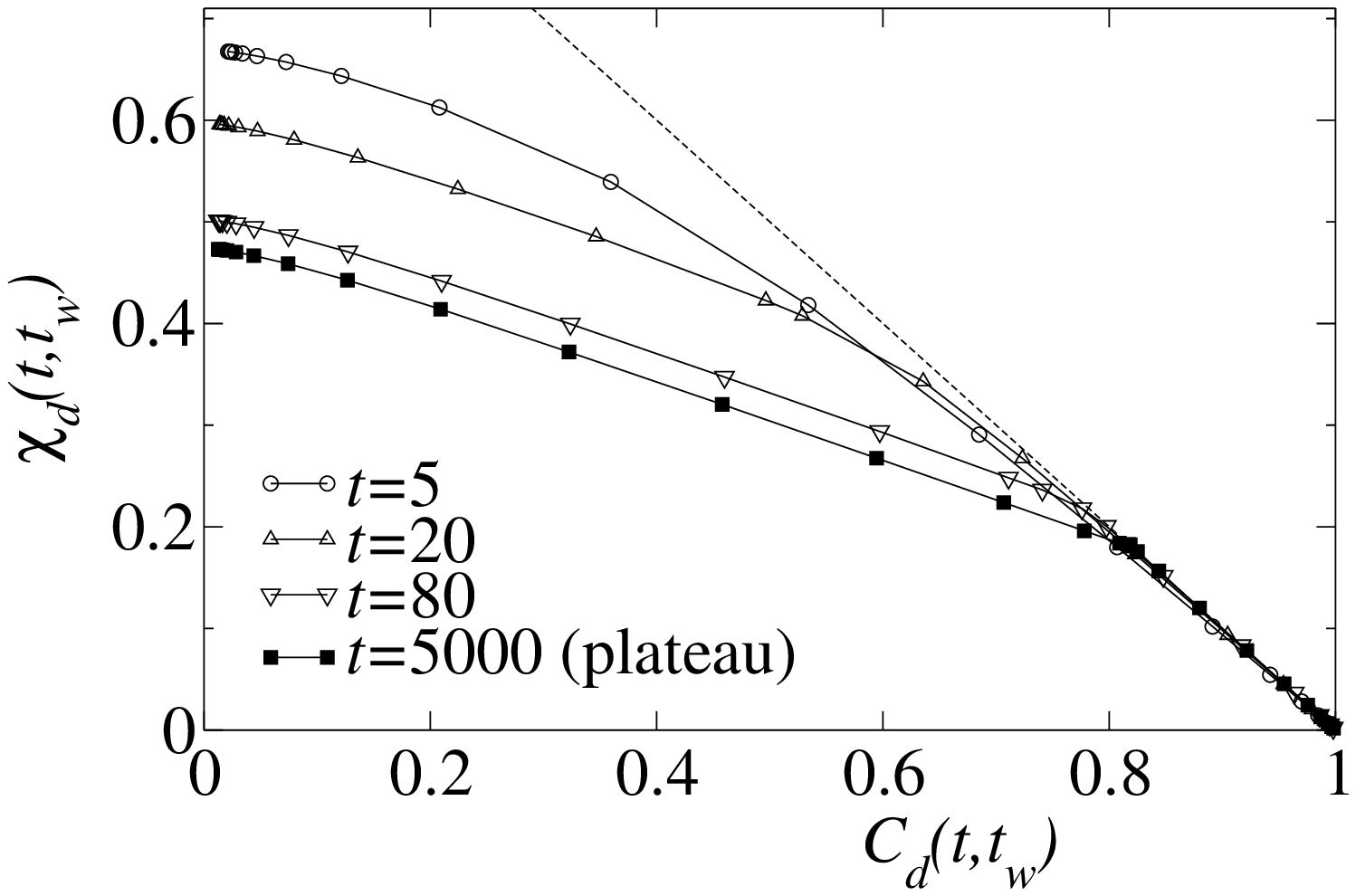,width=0.45\columnwidth}
\epsfig{file=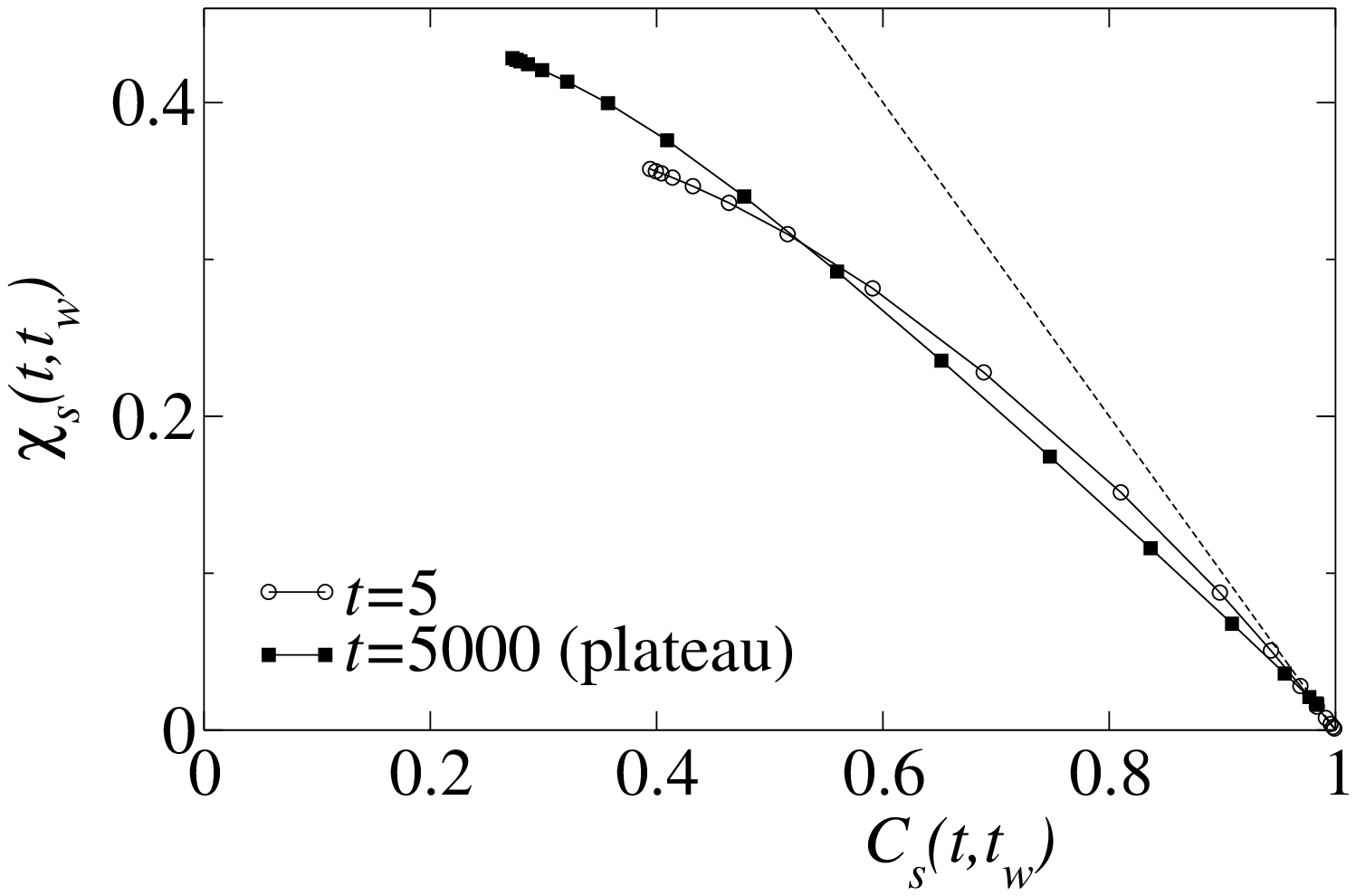,width=0.45\columnwidth}
\caption{FD plots for the first stage of relaxation
in the SPM at $\beta=20$. We show the FDR for $t=5000$, which
is within the plateau of $c(t)$, and for several earlier times, 
at which the system is approaching the plateau.
(Left) Defects, (right) spins. We
show the equilibrium FDT with a dashed line. }
\label{fig:spm_zerotemp}
\end{figure}

We now turn to the
SPM, in which there are two relaxation stages. We begin our
discussion with 
the initial stage, shown in the inset of
Fig.~\ref{fig:tpm_spm}.  The corresponding FD plots for local spin and
defect dynamic functions are shown in Fig.~\ref{fig:spm_zerotemp}.

For the defect variables, we find that the
self overlap within the plateau state is $q_{11}\simeq0.82$.  This
number comes from isolated spins that can flip without energy 
penalty~\cite{BuhGar02}.  The mutual overlap between the plateau and initial
states is very close to zero, $q_{10}\simeq0.01$. 
A well-defined FDR can be determined, $x_{10} \simeq 0.35$, see 
Fig.~\ref{fig:spm_zerotemp}. This makes the FD plot apparently 
very similar to the one found during the aging 
dynamics of mean-field spin glasses characterized by a one-step 
replica symmetry breaking transition
at the static level~\cite{Young,BuhGar02,CugKur}.

For the spin degrees of freedom, we find $q_{11}\simeq0.98$ and 
$x_{11} \simeq 1$. For smaller correlation, however, the 
parametric FD plot is curved and there is no 
well-defined value for $x_{10}$. 

This behaviour is qualitatively similar to that of the FA model,
characterized by an initial ``zero temperature'' relaxation, followed
by a reaction-diffusion aging regime.  However, there are some
important differences. Firstly, the decay into the initial plateau is
exponential in the FA model, and the overlap $q_{10}$ takes its maximal
possible value $q_{10} \simeq \frac{1-n_0}{1-n_{\rm plateau}}$, where
$n_{\rm plateau}$ is the defect density in the plateau state, and
$n_0=1/2$ is the density in the initial state. This reflects the
situation described by Eq.~(\ref{equ:q_East}), in which the decay
proceeds by uncorrelated relaxation events.
Similarly, the FA model has $x_{10} \to 0$ at low
temperatures: this occurs because all Monte Carlo acceptance
probabilities are either zero or unity at $T=0$,
and the dynamics are independent of the perturbing field; thus there
is no response.  On the other hand, the SPM has a finite value for $x_{10}$
even at zero temperature, because there are energy conserving
processes that have field-dependent acceptance probabilities close to $1/2$,
even when $T=0$.

We believe that the non-trivial zero temperature FDRs for both spins and
defects come from the interplay between two types of 
process: reversible `diffusive' spin flips in which the energy
does not change, and irreversible `coagulative' spin flips in which
the energy is reduced.  Both processes are occurring with similar
rates.  As discussed for the TPM, if the reversible process was
dominating we would expect a large value for $x_{10}$, while
irreversible relaxation would lead to a small value of $x_{10}$.

To confirm this analysis and probe in more detail the competition between
diffusive and coagulative processes, 
we modify the Glauber dynamics,
by introducing the additional parameters $\gamma$ in the rates of
Eq.~(\ref{equ:def_W}) (see \ref{app:chat}).  
Our aim is to affect the dynamics of the SPM in such a way that
diffusive and coagulative processes become well-separated 
processes, in order to analyze the consequences for the resulting FD plots.

There are five types of move in the SPM: the
energy change may be any of $\pm2,\pm1,0$.  We choose $\gamma$ to
depend on the modulus of the energy change and label these rates
according to the number of defects adjacent to the spin that
flips, $u$. This number satisfies $u_\mathrm{final} = 4-u_\mathrm{initial}$,
so we define three $\gamma_{u_\mathrm{initial}}$ factors for the three
possible cases:
\begin{eqnarray}
(u=4) \leftrightarrow (u=0) & \qquad \hbox{multiplier} \;& \gamma_4; \nonumber\\
(u=3) \leftrightarrow (u=1) & \qquad \hbox{multiplier} \;& \gamma_3; \nonumber\\
(u=2) \leftrightarrow (u=2) & \qquad \hbox{multiplier} \;& \gamma_2. \nonumber
\end{eqnarray}
The multipliers are the same for both forward and reverse
processes, since the dynamics respect detailed balance.  At zero
temperature, the rates for the three forward processes are all
independent and are given by $(\gamma_4,\gamma_3,(\gamma_2/2))$.  We
find that the behaviour depends only weakly on $\gamma_4$ as long as
$\gamma_3 \geq \gamma_4$. We therefore set
$\gamma_3=\gamma_4=1$, so that $\gamma_2$ is a dimensionless
measure of the rate for the energy-conserving, diffusive, events.

\begin{figure}\hfill
\epsfig{file=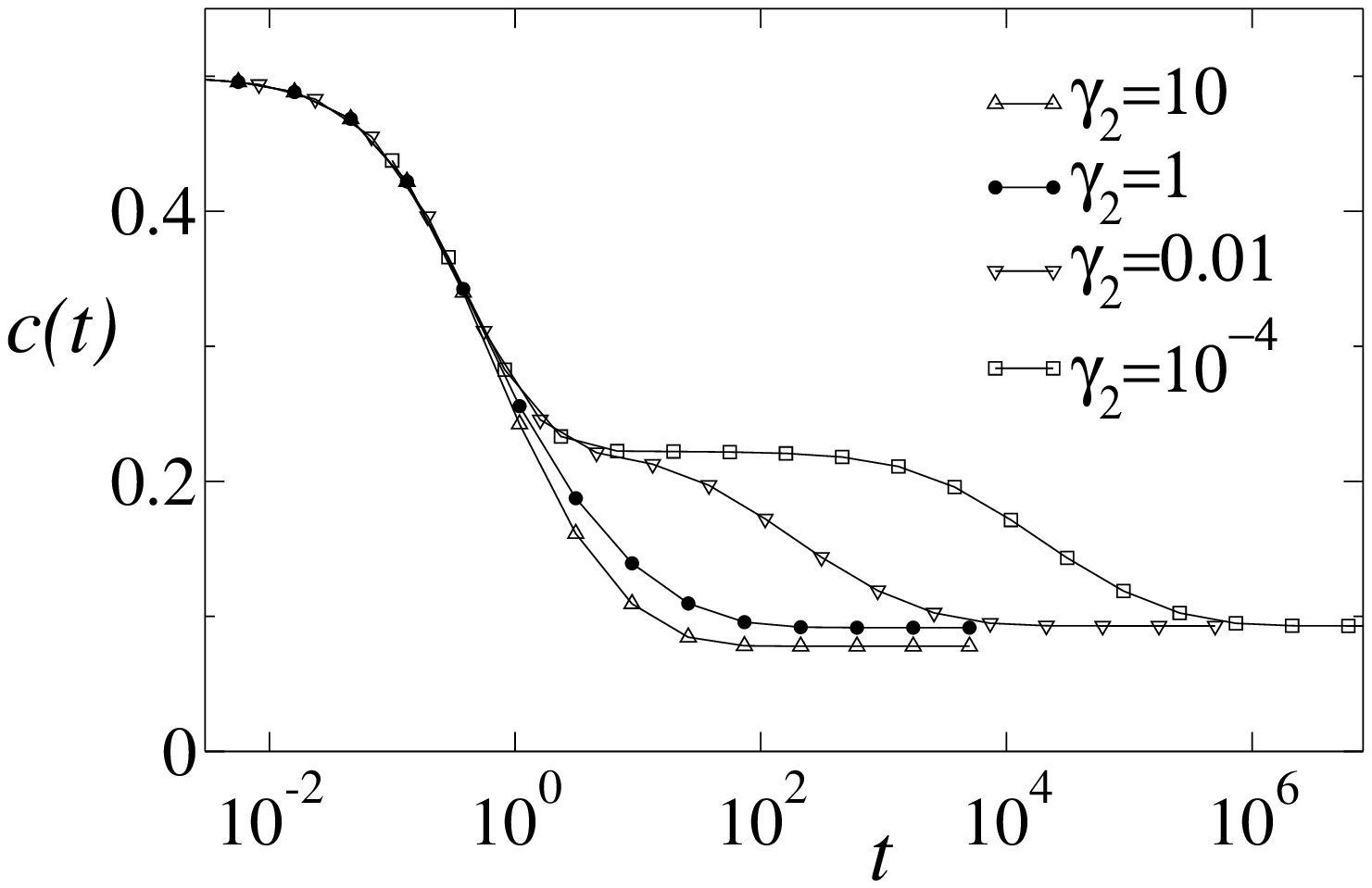,width=0.45\columnwidth}
\epsfig{file=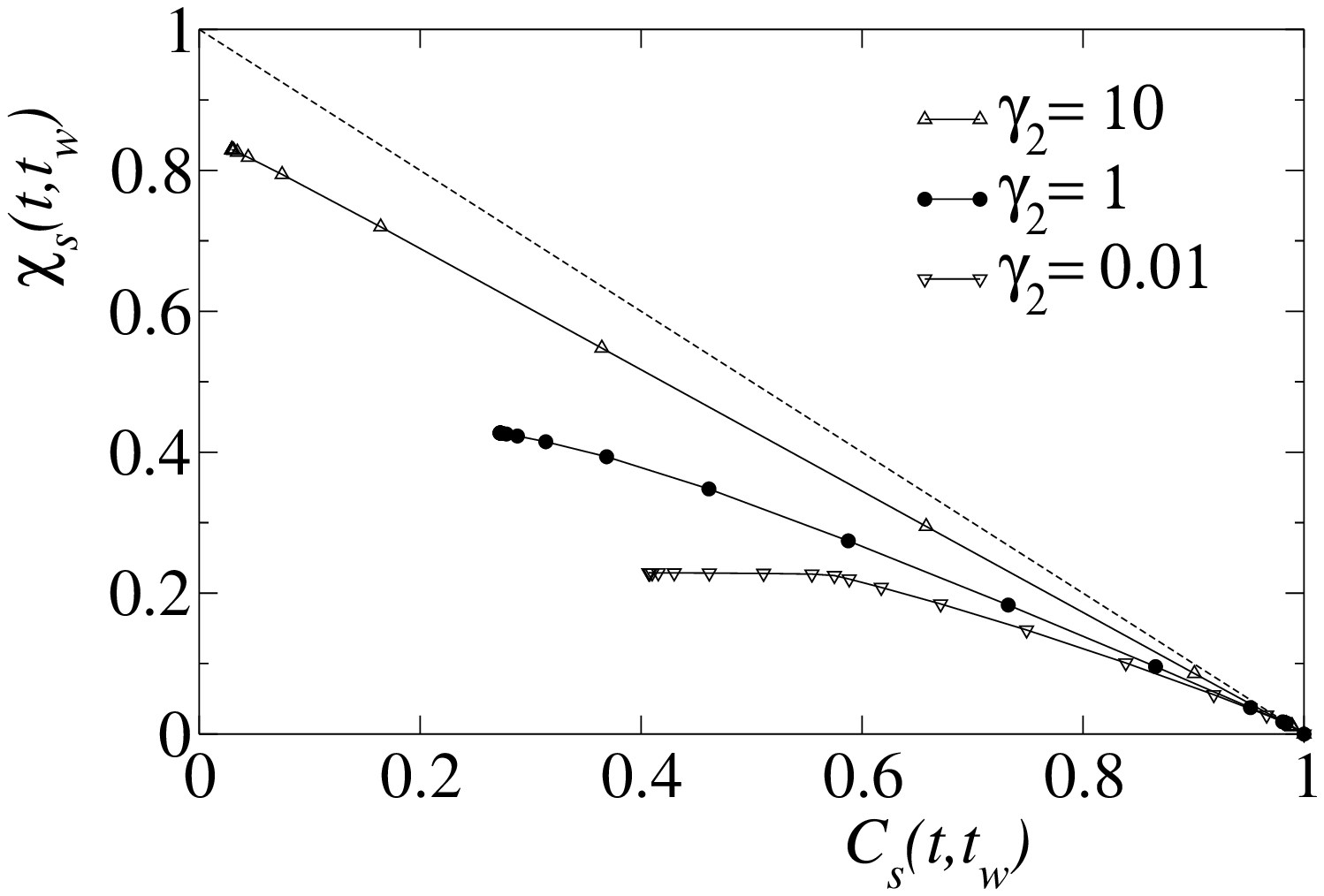,width=0.45\columnwidth}
\caption{ 
Correlation and response at $T=0$, and their
dependence on $\gamma_2$.
(Left)~Energy density $c(t)$ (linear scale) after a quench to $T=0$.
Two stages are clearly visible at small $\gamma_2$; at large $\gamma_2$,
the extra energy-conserving processes allow more efficient exploration
of configuration space, and the energy relaxes to a smaller value.
(Right)~Parametric plot for spin observables at the latest values of $t$.
At small $\gamma_2$, the first stage of relaxation
is purely `coagulative' so it has $x\simeq0$; the second stage has
both `diffusive' and `coagulative' processes, so $x$ is finite. 
At large $\gamma_2$, energy-conserving (diffusive) processes dominate, 
and the FDR is large. 
}
\label{fig:spm_gamma}
\end{figure}

We show the energy decay and corresponding FD plots 
for spin observables in Fig.~\ref{fig:spm_gamma}, 
as a function of $\gamma_2$.  
For $\gamma_2\ll1$, the zero temperature
relaxation takes place in two stages, unlike the case
of Fig.~\ref{fig:tpm_spm}, in which $\gamma_2=1$. The first
stage corresponds to pure coagulation: the system
takes the path of steepest descent in the energy 
landscape. This process has an FDR of zero,
as was the case for the first stage of relaxation in the TPM. 
In the second stage, the SPM explores its energy landscape 
by decreasing the energy where possible;
otherwise, it makes energy-conserving moves.
This stage has a finite FDR
that reflects the ratio of energy-conserving, diffusive events (along
flat directions of the energy landscape) and coagulative events
(along downward directions).  We also note that this second stage is not
a simple exponential decay, so the structure
of Fig.~\ref{fig:spm_zerotemp} is not simply a result of two
exponential relaxation mechanisms.

In the case $\gamma_2\gg1$, the increased rate for diffusive
moves means that the system prefers to explore the energy landscape
in directions along which the landscape is flat. It therefore
explores more of configuration space, and is more
likely to find pathways to low energy.  Thus, the plateau
state contains fewer defects than that for $\gamma_2=1$.
Since the diffusive moves explore the constant energy
surface in an unbiased way, the system tends to equilibrate
locally between each coagulative move, and the FDR approaches
unity as $\gamma_2$ is increased.

Overall, we see that the finite value of $x_{10}$ observed
in Fig.~\ref{fig:spm_zerotemp} comes from competition between diffusive
and coagulative processes.  
Its value, $x_{10} \simeq 0.35$ for $\gamma_2=1$ 
in fact depends continuously on the microscopic rates in
the problem: we conclude that it is a non-universal number that
measures the extent to which coagulation dominates over diffusion
in the relaxational dynamics. In  particular, the FDR 
has no connection with static quantities, in contrast to
mean-field models~\cite{CugKur,FraMezParPel98}.

Moving to spatial structure, 
Eqs.~(\ref{equ:tpm_twospin:c}, \ref{equ:tpm_twospin}) 
also hold for the SPM, and spin
correlation and response functions are independent of wave vector $k$.  
In the short time regime, the defect response is $\mathcal{O}(k)$ at
small $k$. This is because the zero temperature response comes only from
energy-conserving processes,
and the rates for these processes couple only
to temperature gradients, and not to the absolute temperature.  
Our numerical simulations confirm that the response vanishes at
small $k$. This was already demonstrated for the 
TPM in Fig.~\ref{fig:tpm_cq} so we do not show more data for the SPM.

\subsection{Aging regime: activated dynamics}

\begin{figure}\hfill
\epsfig{file=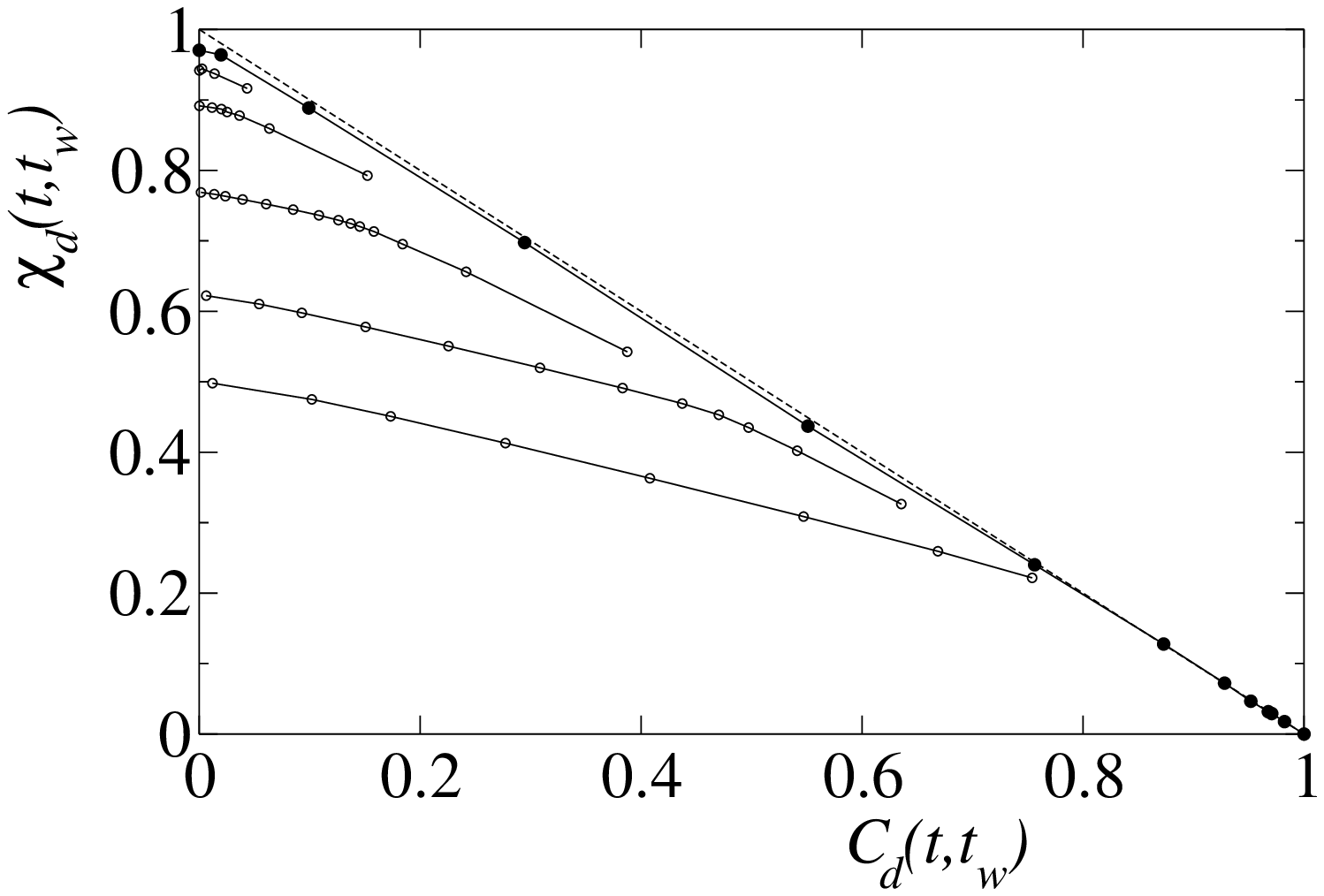,width=0.45\columnwidth}
\epsfig{file=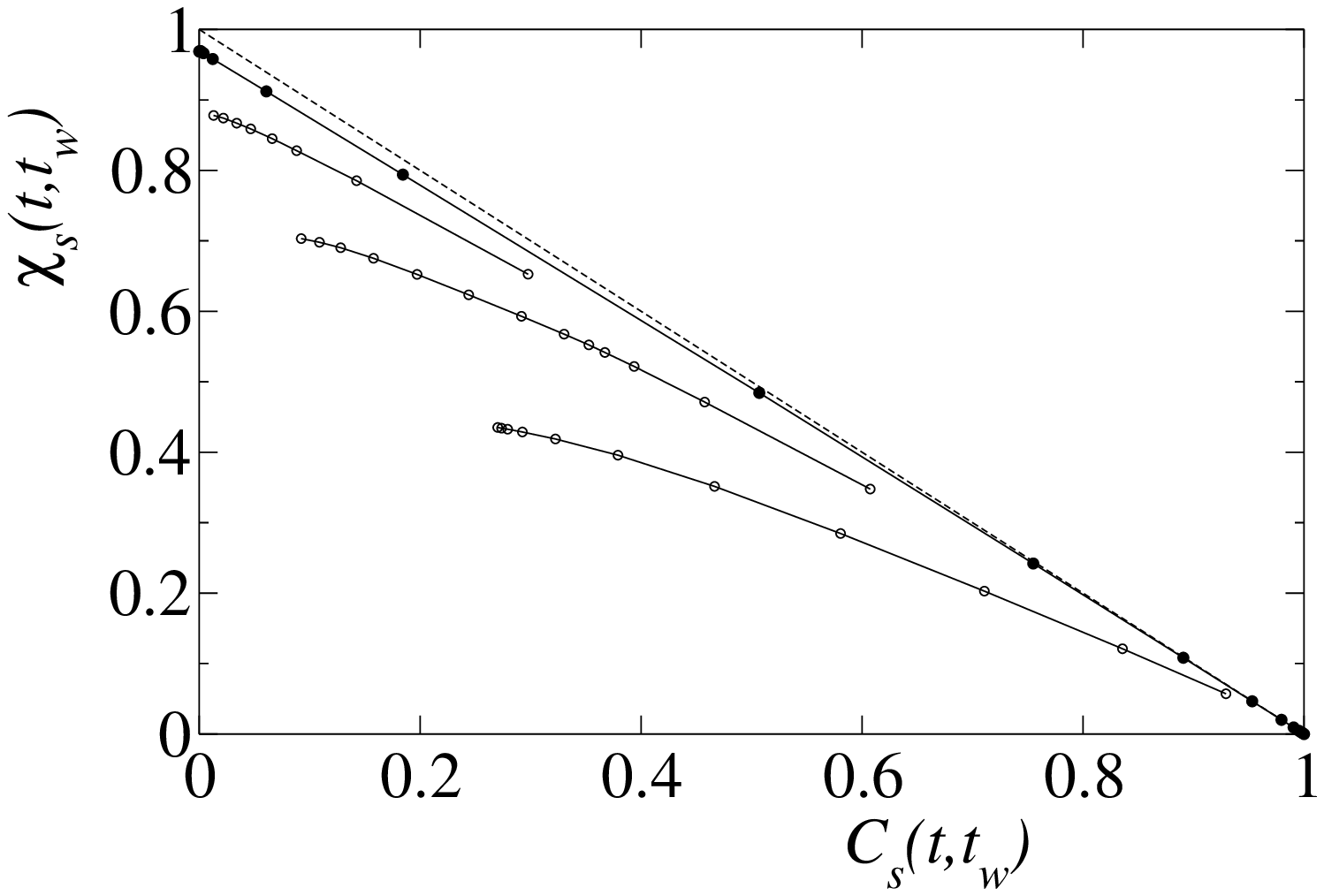,width=0.45\columnwidth}
\caption{Correlation and response in the aging regime for the SPM. 
(Left) Defects: we show
$t\in\{140,2500,1.0\times10^4,4.3\times10^4,1.8\times10^5,7.4\times10^5\}$,
which decrease from top to bottom. 
(Right) Spins: final times are $t\in\{140,1.6\times10^4,
1.1\times10^5,7.4\times10^5\}$, with the same trend. For both cases, the 
first time is representative
of the initial plateau in $c(t)$: recall Fig.~\ref{fig:spm_zerotemp}.
Except for the latest time, we 
concentrate on the region in which equilibrium FDT is not obeyed;
the parametric plot goes through $(C,\chi)=(1,0)$, for
all values of $t$, but we do not plot data in this region.
}
\label{fig:spm_age}
\end{figure}

When the SPM is quenched to a small but finite temperature, the system 
eventually leaves the jammed
state into which it relaxed initially, and enters an aging regime, in
which the energy decays as a power law.  We show local correlation-response 
data for both spins and defects in
Fig.~\ref{fig:spm_age}.  As $t$ and $\tw$ get large, the local FDR
approaches unity for all values of $C$, even though the system is
still out of equilibrium.  

In the framework of the previous sections, the approach of the FDR
to unity comes from the fact that defects are very sparse at large $\tw$,
and coagulation events become
increasingly rare. The result is that diffusive processes
dominate the response, and lead to $X(q,t)\simeq1$, for finite $q$ and 
large $t$ [recall that $X(q,t)$ is the FDR, evaluated at $C(t,\tw)=q$; see
(\ref{equ:X_qt})].  However, it is now a familiar feature of reaction-diffusion
systems that there is structure in $X(q,t)$ for finite $t$ and small $q$
\cite{MayBerGarSol,Mayer06}.
We define 
\begin{equation}
X_\infty(t) =  \lim_{q\to0} X(q,t).
\end{equation}
It was recently shown \cite{Mayer06}
that in the aging regime of the FA model, $X_\infty(t)$ is
a negative number that depends on the dimension of the system,
but not on the time $t$.  Observables associated with domain walls
in the Glauber-Ising chain have similar
behaviour, with $X_\infty=0$ \cite{MayBerGarSol}.
However, for both these models, 
$X(q,t)$ only converges to its limiting
value at very small $q$. This requirement also necessitates working
at large $t$: the limit of long times and very small $q$ makes 
direct investigation of this limit very difficult in simulations. 
For the FA model and the Glauber-Ising chain, 
$X_\infty$ is independent of the observation wave vector, $k$. 
From the point of view of simulations, it is fortunate that  
$X_\infty$ is typically
much easier to measure at small wave vector. 

\begin{figure}\hspace{72pt}
\epsfig{file=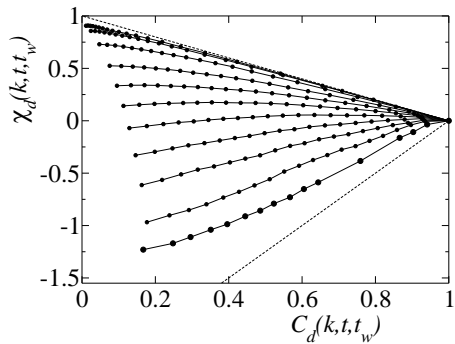,width=0.45\columnwidth}
\caption{Parametric FD plot for spatially dependent defect
observables at $\beta=5$ and $t=7.4\times10^4$ (this time is
within the aging regime). The wave vectors are $k/k_\mathrm{max}\in\{
1.0,0.50,0.33,0.24,0.20,0.17,0.14,0.11,0.083,0.050,0.0\}$,
with $k$ decreasing from top to bottom. Dashed lines show the
equilibrium FD relation, and the small $k$ prediction of 
(\ref{equ:X_field}). The deviations from the small $k$ prediction
arise from corrections to the aging limit, as discussed
in the main text and in Fig.~\ref{fig:spm_age_k0}.
The response at $k=0$ was calculated using an explicit 
field, rather than the no-field method described in 
\ref{app:chat}, 
since that method is rather inefficient for responses at small $k$.
}
\label{fig:spm_age_k}
\end{figure}

In Fig.~\ref{fig:spm_age_k}, we show that the spatial structure of
defect correlation and response in the
SPM is consistent with this behaviour. That is,
$X(q,t)$ is positive and close to unity for large $q$, and negative
for small $q$.  The value of $q$ at which the crossover takes place depends 
on the observation wave vector $k$; its value is close to unity at large
$k$ and close to zero at small $k$. 
The physical origin of a negative asymptotic FDR is the same as for the 
FA and East models~\cite{Mayer06}: when applying a small temperature 
change $\delta T >0$, ($T$ is the field conjugate to the energy), 
the dynamics of the system gets accelerated 
and the energy decays faster towards
equilibrium so that $\delta c(t) <0$, and indeed $\delta c(t) / 
\delta T <0$, unlike the equilibrium case.

We now discuss how the limiting negative value of the FDR 
and the overall behaviour 
observed in fig.~\ref{fig:spm_age_k} can be predicted by a similar
calculation to that used for the FA model in Ref.~\cite{Mayer06}.  
The calculation for the SPM is slightly more involved,
due to the presence of two growing length scales in the aging
regime of that model. As discussed in \cite{Jack05-spm},
the reaction-diffusion dynamics involve exchange of defect
dimers between isolated single defects. This process takes
place over a length scale $\ell(t)\sim [1/n_i(t)]$; at 
long times then this length scale is much larger than
the correlation length for density fluctuations in the aging
regime, which is $\xi(t)\sim [1/n_i(t)]^{1/2}$. The processes
by which dimers are exchanged have an energy barrier of two.
Their rates are also inversely proportional to the distance
$\ell(t)$
over which the dimer must travel \cite{BuhGar02,Jack05-spm}.
The result is that these rates depend linearly on the 
defect density in the system, averaged over a region of
linear extent $\ell(t)$.

In \ref{app:field}, we show how this situation can be described
by an effective reaction-diffusion system, where correlations
in the aging regime
can be calculated in a field-theoretic formalism. The
rates for reaction and diffusion depend the density,
averaged over a region of fixed size $\ell$.
We neglect the time dependence of this
length scale, but our results depend
only on whether $k\ell$ is large or small compared to unity. 
Hence, our results apply if $k\ell(\tw)$ and $k\ell(t)$
are both small, or both large. The other parameter that enters
the FDR is $z_k(\tw)=Dk^2/[2\lambda n(\tw)]$, which measures
the relative size of $k$ and $\xi(t)$. 

The calculation described in \ref{app:field} is
quite lengthy, but the important result 
is (\ref{equ:X_field}). At small $k$, the FD plot is a straight
line with gradient $X=-(5/2)$. At large $k$, we expect 
quasi-equilibrium behaviour, $X=1$ (reversible `diffusive' 
processes dominate on small length scales). Further, we can
identify an intermediate regime, in which $k\ell(t)$ and $k\ell(\tw)$
are both large, but $k\xi(t)$ and $k\xi(\tw)$ are both small.
We find that this regime has $X=-3$. 

Comparing with the results of figure~\ref{fig:spm_age_k},
the behaviour is in qualitative agreement with the field
theoretic calculation. However, quantitative agreement
is rather poor for the limit of small wave vector,
and the intermediate regime is not apparent. 
To understand this,
we observe that the field theoretic calculation is valid
only deep in the aging regime, which requires
\begin{eqnarray}
\langle n_i(\tw)\rangle, \langle n_i(t) \rangle \ll 1, \qquad 
\langle n_i(t) \rangle \gg e^{-\beta}
\label{equ:field_bounds}
\end{eqnarray}
We also predict logarithmic corrections to these
tree level results, since the critical dimension
of the reaction-diffusion system is $d_c=2$ (to show this,
one must be careful to take $n(t) \ell(t)^2$ to infinity, in the 
limit of large time).

\begin{figure}\hfill
\epsfig{file=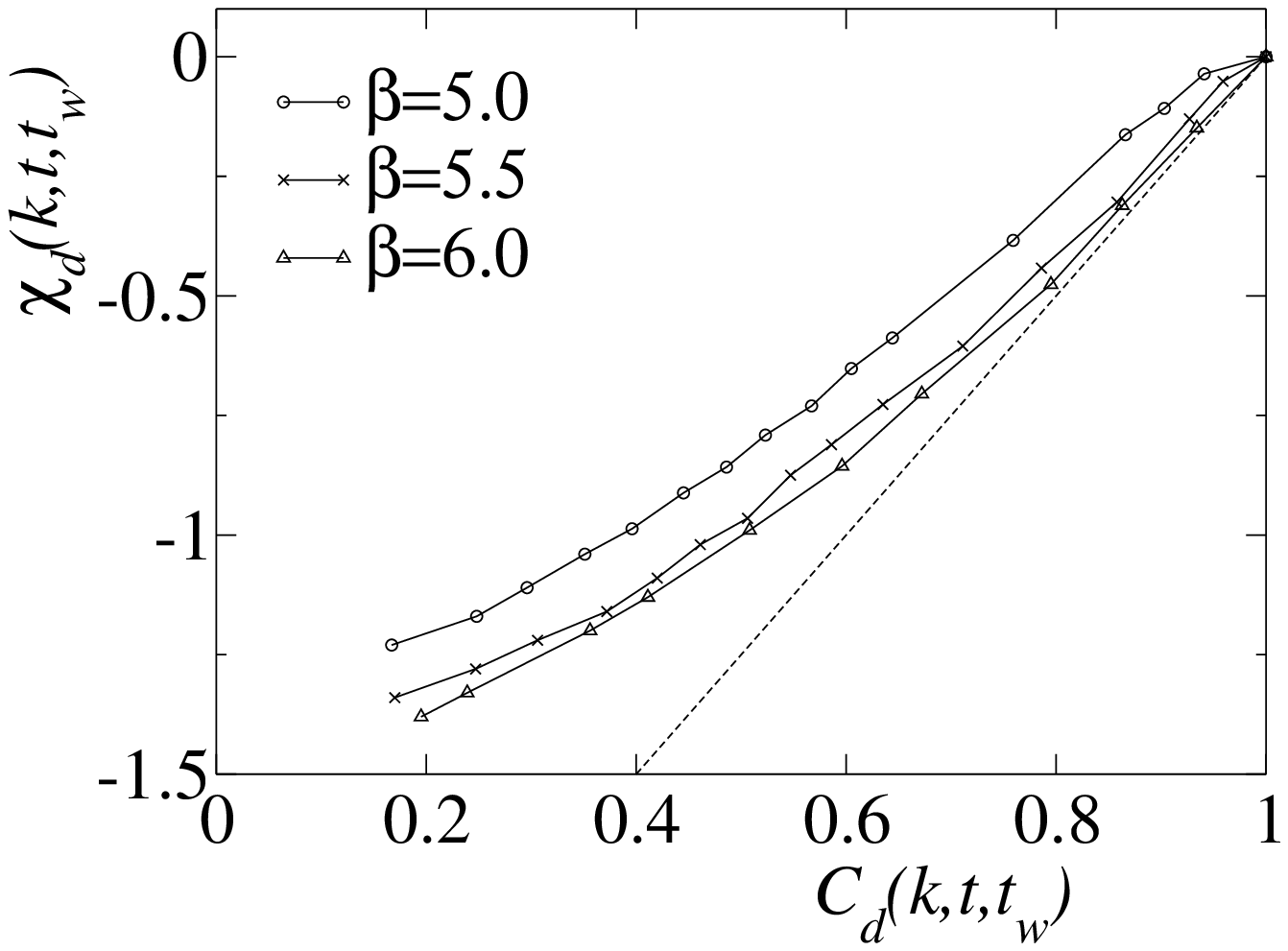,width=0.45\columnwidth}
\epsfig{file=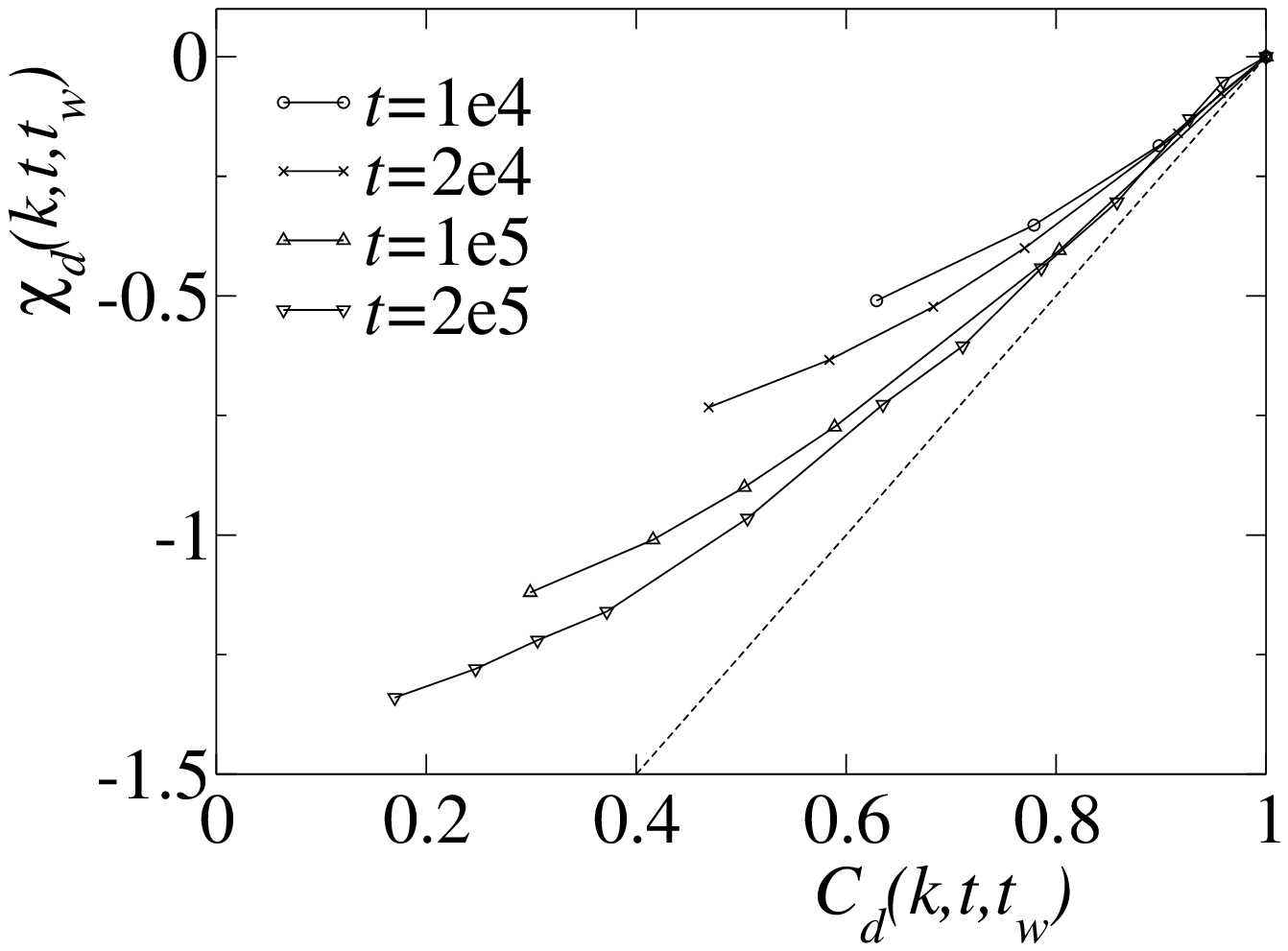,width=0.45\columnwidth}
\caption{
Parametric FD plots in the SPM at $k=0$. As the bounds of 
(\ref{equ:field_bounds}) are improved, the
data approaches the prediction $X=-(5/2)$,
which is shown as a dashed line.
(Left) Increasing $\beta$ with a constant
value of the rescaled time ($c^2t=3.3$) improves the bound
$\langle n_i(t) \rangle \gg e^{-\beta}$. This improves
the agreement with theory
for $\tw$ close to $t$.
(Right) Increasing $t$ and $\tw$ at fixed 
temperature improves the bound
$\langle n_i(\tw) \rangle \ll 1$,
which increases the range over which the system
is close to the prediction for the aging limit.
}
\label{fig:spm_age_k0}
\end{figure}

The conditions of (\ref{equ:field_bounds}) 
are rather restrictive,
necessitating simulations at low temperatures and long times.
While we were not able to access the regime of
intermediate $k$, we show in
figure~\ref{fig:spm_age_k0} that the FDR at $k=0$ does 
appear to approach $-(5/2)$ as we improve the bounds of 
(\ref{equ:field_bounds}). Our data is therefore consistent
with a value of $X_\infty=-(5/2)$, although
convergence to this limit is rather slow. 

\section{Conclusion}

Finally, we bring together the results of the previous sections and
discuss their significance. We have shown that the FDR in
these plaquette models has considerable structure. 
By using the `no-field' method of Refs.~\cite{Chatelain04,Ricci}, 
we have been able to perform 
systematic measurements of correlation-response 
relations allowing resolution of the questions about
the SPM and TPM that were raised in Refs.~\cite{BuhGar02,Garrahan-Newman}.

By definition,
the FDR measures the ability of a system to respond to an external 
perturbation (normalised by an equilibrium system with the same
correlations). We have argued throughout that when the temperature is small,
the system can only respond by `diffusive' energy-conserving processes.
A large FDR means that these processes are dominating (for example,
this occurs in dilute reaction-diffusion systems); a small FDR means that
irreversible `coagulative' processes are dominating (for example, the
East model). 
This is consistent with our interpretation of the FDR as a
purely dynamic object in these systems, which have a single pure
state at all finite temperatures.
It contrasts with the situation following
a quench into a regime in which ergodicity is broken, where the
long time limit of the FDR should reflect the underlying pure state
structure~\cite{CugKur,FraMezParPel98}.

We have shown in particular that the multi-stage 
relaxation of the plaquette models leads to well-defined 
FDRs within each stage, with values ranging from $X=1$ for 
quasi-equilibrium regimes, non-trivial $0<X<1$ for local observables 
within activated stages, and non-trivial $X<0$ for global observables
coupled to thermal activation. Therefore, plaquette models 
present an extremely rich aging behaviour that should be 
considered as an alternative paradigm to mean-field models 
when interpreting numerical data on physical systems such as 
disordered magnets or supercooled liquids.

\begin{ack}

We thank Peter Mayer and Peter Sollich for useful discussions.
This work was supported by NSF grant CHE-0543158 (RLJ); by
EPSRC grants GR/R83712/01 (RLJ and JPG) and
GR/S54074/01 (JPG); and by University of Nottingham grant no.\ FEF 3024 
(JPG).

\end{ack}

\appendix

\section{Measuring response functions}
\label{app:chat}

In this Appendix we give the generalisation of the `no-field'
method of \cite{Chatelain04,Ricci} to continuous time
processes: we used this method to calculate all responses to
local fields in this article.

Suppose that we have a spin
system with Hamiltonian $H=H_0-\sum_i h_i s_i$, where
$s_i\in\{-1,1\}$; the sum is over all the spins,
and the $h_i$ are time-dependent local fields.
We choose a continuous time dynamics that obeys detailed balance,
in which the spin on site $i$ flips with a rate
\begin{equation}
W_i = \frac{\gamma_i}{1+\exp[\beta(\Delta_i+2h_is_i)]},
\label{equ:def_W}
\end{equation}
where $\beta$ is the inverse temperature; $\Delta_i$ is the 
change in $H_0$ on flipping spin $i$; 
and $\gamma_i$ is a local rate that does not
depend on the state of spin $i$ or on the field $h_i$.  Glauber
dynamics is the case when $\gamma_i=1$ throughout. In kinetically
constrained models, we have $\gamma_i\in\{0,1\}$ according to
state of the neighbours of spin $i$. In section~\ref{sec:spm_init}, 
we consider a modified
Glauber dynamics, such that $\gamma_i$ depends on the environment of
each spin. However, the local field $h_i$ always couples to the
dynamics as in (\ref{equ:def_W}).

The application the method of \cite{Chatelain04,Ricci}
to this continuous time system is as follows:
the integrated response of observable $A$ to the field $h_i$ is
\begin{equation}
\chi^{(A)}_i(t,\tw) \equiv \left.\frac{\mathrm{d} \langle A(t)
\rangle}{\mathrm{d}(\beta h_i)}\right|_{h_i=0} = \langle A(t)
\mathcal{R}_i(t,\tw) \rangle_{h_i=0}
\end{equation}
where the field $h_i$ acts on site $i$ between the times $\tw$ and $t$,
and 
\begin{equation}
\mathcal{R}_i(t,\tw) 
 =  
\sum_{\mathrm{flips} \in (\tw,t)} \frac{-2s_{i,t_\mathrm{flip}^-}}
{1+\exp(-\beta\Delta_{i,t_\mathrm{flip}^-})}
+ \int_{\tw}^t \mathrm{d}t'\, \frac{2s_{it'} \gamma_{it'}
\exp(\beta\Delta_{it'})}{[1+\exp(\beta\Delta_{it'})]^2},
\label{equ:chat_response}
\end{equation}
where the sum is over the flips of spin $i$ within
the given time window; 
the notation $t_\mathrm{flip}^-$ indicates that the
summand is to be evaluated just before the spin flip. The strength
of this method is that the small field $h_i$ does not enter
(\ref{equ:chat_response}).

We now outline a derivation of (\ref{equ:chat_response}).  Let a
trajectory be a sequence of configurations of the system at a series
of time steps, where sequential configurations differ in at most one
spin. The probability of a given trajectory is given by
\begin{equation}
P_\mathrm{traj} = \left[
\prod_{\tau=1}^{t/\epsilon} W_{\st_\tau,\st_{\tau-1}}
\right]
p_0(\st_0),
\end{equation}
where $\st_\tau$ represents the state of the whole system in the
$\tau$th configuration; $p_0(\st_0)$ is the probability of the initial
condition; $\epsilon$ is the microscopic time step; and
\begin{equation}
W_{\st',\st} = \delta_{\st',\st} 
\left( 1- \epsilon \sum_i W_i \right)
 + \epsilon \sum_i \delta_{s_i+s_i'} W_i
\end{equation}
is the probability of a transition from state $\st$ to state $\st'$,
in a time $\epsilon$. We
use $s_i$ (or $s_i'$) to denote the state of spin $i$ in configuration $\st$
(or $\st'$); the rates $W_i$ are defined in (\ref{equ:def_W}). 

We note that the microscopic time step must be small enough to ensure
that $W_{\st',\st}$ is always positive. Simple discrete time Monte Carlo
dynamics have $W_i\leq 1$ for all $i$, and $\epsilon=N^{-1}$ where $N$
is the total number of spins.  Here we explicitly allow for $\epsilon
< N^{-1}$ since we wish to take the continuous time limit. Also,
recall that $P_\mathrm{traj}$ is defined only for trajectories in
which successive configurations differ in at most one spin. Thus we
have $\sum_{\st'} W_{\st'\st} = 1$, as long as the sum is taken over states
$\st'$ that differ from $\st$ in at most spin; this ensures conservation
of probability.

The definition of the stochastic average is then
\begin{equation}
\langle A \rangle = \sum_\mathrm{traj} A_\mathrm{traj} P_\mathrm{traj},
\end{equation}
where $A_\mathrm{traj}$ is the value of the observable $A$ for the
given trajectory, and the sum is over all possible trajectories.

To calculate the integrated response, we simply write
\begin{eqnarray}
\frac{\mathrm{d}}{\mathrm{d}(\beta h_i)} P_\mathrm{traj} &=& 
\sum_{\tau'=(\tw/\epsilon)+1}^{t/\epsilon}
\left[\prod_{\tau=\tau'+1}^{t/\epsilon} W_{\st_\tau,\st_{\tau-1}}
\right] \times 
\nonumber \\ & &
 \left[\frac{\mathrm{d}}{\mathrm{d}(\beta h_i)} W_{\st_{\tau'},\st_{\tau'-1}}
\right] \left[
\prod_{\tau=1}^{\tau'-1} W_{\st_\tau,\st_{\tau-1}}\right]p_0(\st_0)
,
\end{eqnarray}
where we assumed that the field acted only between times $\tw$ and
$t$, as before. We then write $[\mathrm{d} W_{\st'\st} /\mathrm{d}(\beta
h_i) ] = R_{\st'\st} W_{\st'\st}$, which defines the matrix elements 
$R_{\st'\st}$ (at least for $W_{\st'\st}>0$, which is the only relevant 
case). Hence,
\begin{equation}
\frac{\mathrm{d}}{\mathrm{d}(\beta h_i)} \langle A \rangle =
\sum_\mathrm{traj} A_\mathrm{traj} P_\mathrm{traj}
\sum_{\tau=\tw/\epsilon}^{t/\epsilon} R_{\st_\tau,\st_{\tau-1}}
.
\end{equation} 
Explicitly constructing the matrix $R_{\st_\tau,\st_{\tau-1}}$ as in
Ref.~\cite{Chatelain04}, and then taking the limit of continuous time,
$\epsilon\to0$, leads to (\ref{equ:chat_response}).

To calculate the defect response in the plaquette models,
we also require the response for cases where the random field couples
not to a single spin, but to a local function of several spins. For
example, consider the response to a perturbation
$$
\delta H = -g_{i} \, s_1 s_2 s_3 s_4 \equiv 
-g_{i} n_{i}
$$ where the second equality defines 
the dual plaquette variable $n_{i}$. In that case,
\begin{equation}
\frac{\mathrm{d}}{\mathrm{d} (\beta g_{i})} \langle A(t) \rangle
 = \langle A(t) \mathcal{S}_{i}(t,\tw) \rangle
\end{equation}
where the perturbation acts between times $\tw$ and $t$ as before, and
\begin{eqnarray}
\mathcal{S}_{i}(t,\tw) &= \sum_{a\in \{1,2,3,4\}} &\Bigg\{
\sum_{\mathrm{flips} \in (\tw,t)}
\frac{-2n_{i,t_\mathrm{flip}^-}}
{1+\exp(-\beta\Delta_{a,t_\mathrm{flip}^-})} 
\nonumber \\ & &  +
\int_{\tw}^t \mathrm{d}t'\, \frac{2n_{i,t'} \gamma_{at'}
\exp(\beta\Delta_{at'})}{[1+\exp(\beta\Delta_{at'})]^2} \Bigg\}.
\end{eqnarray}
The sum inside the curly brackets is over flips of spin $a$,
and the summand is again evaluated just before the spin flip.

\section{Field theory for defects in the SPM}
\label{app:field}

In this appendix, we define an effective reaction-diffusion
system that takes account of the two growing length scales
in the aging regime of the SPM. We calculate that
correlation and response in this system at tree-level,
following \cite{Lee94}. The methods are quite standard,
so we quote only the main results; the review of \cite{tauber} 
gives technical details of this formalism.

\subsection{Effective theory and dynamical action}

The effective theory is defined for defects on a cubic lattice
of $N$ sites
in $d$ dimensions, with integer occupancy on each site. 
Diffusion 
takes place between neighbouring sites, and annihilation takes
place on a single site. That is,
\begin{equation}
  \begin{array}{cccl}
n_i n_j &\to& (n_i-1), (n_j+1), & \quad  \hbox{rate } n_i D_{ij} \\
n_i &\to& (n_i-2), & \quad \hbox{rate } n_i (n_i-1) \lambda_i.
  \end{array}
\end{equation}
The rates $D_{ij}$ and $\lambda_i$ depend on the local defect density.
We also require their coupling to perturbations to the energy
and to the inverse temperature $\beta$. We write the energy as 
$E=\sum_i n_i - \sum_i h_i n_i$, and the appropriate rates
are
\begin{eqnarray}
D_{ij} &=& 2D_0 \frac{ e^{-\beta(2-h_i-h_j)} }{ 1+e^{\beta(h_i-h_j)} } 
\sum_{i'} g_{i';ij}^{(D)} n_{i'} \\
\lambda_{i} &=& \lambda_0 e^{-2\beta(1-h_i)} \sum_{i'} g_{i';i}^{(\lambda)} 
n_{i'}.
\end{eqnarray}
Both rates are of activated form, with energy barriers close to two, since
they require an intermediate stage containing a dimer.
The factor $e^{\beta(h_i-h_j)}$ in the denominator
of $D_{ij}$ sets the coupling to the random potential $h$. The
functions $g_{i';ij}^{(D)}$ and $g_{i';i}^{(\lambda)}$ decay with
the distance between sites $i$ and $i'$, on a length scale $\len$. They
vanish at $i'=i$ and $i'=j$, and $g_{i';ij}^{(D)}$ is symmetric in 
$i$ and $j$. We also normalise such that 
$\sum_{i'} g_{i';ij}^{(D)} = \sum_{i'} g_{i';i}^{(\lambda)}=1$.

The master equation for this system can be written in an operator
formalism \cite{Doi-Peliti}. We introduce
a bosonic algebra: $[a_i,a_j]=[a_i^\dag,a_j^\dag]=0$; 
$[a_i,a_j^\dag]=\delta_{ij}$, and a vacuum state $|0\rangle$,
such that
$a_i|0\rangle=0$, for all $i$. Let the probability of a
configuration $\{n_i\}$ be $P(\{n_i\})$, and  let
$|P(t)\rangle=\sum_{\{n_i\}} P(\{n_i\},t) \prod_i (a^\dag_i)^{n_i}
|0\rangle$.
Then the master equation is
\begin{equation}
\partial_t |P(t)\rangle = - \hat{W} |P(t)\rangle
\end{equation}
with
\begin{eqnarray}
\hat{W}&=&\sum_{\langle ij\rangle} (1/2)(a^\dag_i-a^\dag_j)
[ (\hat{D}_{ij} + \hat{D}_{ji})(a_i-a_j) 
+ (\hat{D}_{ij}-\hat{D}_{ji}) (a_i+a_j) ] \nonumber \\ &&
+ \sum_i \hat{\lambda}_i [(a^\dag_i)^2-1] a_i^2 
\end{eqnarray}
where $\hat{D}_{ij}$ and $\hat{\lambda}_i$ are now operators,
since they depend on the densities $n_{i'}=a^\dag_{i'} a_{i'}$.

We now construct a path integral representation of the dynamics.
We work in the coherent state
representation, in which operators have matrix elements
$O(\phi,\bar\phi)=\langle 0| e^{\sum_i(\bar\phi_i+1)a_i}
\hat{O} e^{\sum_i\phi_i a_i^\dag} | 0\rangle e^{-\sum_i 
(\bar\phi_i+1)\phi_i}$. 
We take the continuum limit, denoting the lattice
spacing by $l_0$, and the position of site $i$ by $\bm{r}_i$;
we define
$\phi(\bm{r})=\sum_i \phi_i\delta(\bm{r}-\bm{r}_i)$ and
$\bar\phi(\bm{r})=(l_0)^d \sum_i \bar\phi_i\delta(\bm{r}-\bm{r}_i)$,
so that the dimensions of the field
$\phi(\bm{r},t)$ are those of density, and $\bar\phi(\bm{r},t)$ is
dimensionless.  In a similar way,
the field $h(\bm{r},t)=(l_0)^d \sum_i h_i\delta(\bm{r}-\bm{r}_i)$
has dimensions of energy.
The local density of excitations is
\begin{equation}
\rho(\bm{r},t)=[1+\bar\phi(\bm{r},t)]\phi(\bm{r},t).
\end{equation}
We also define $g(\bm{r})=g_{i';i} \delta[\bm{r}-(\bm{r}_{i'}-\bm{r}_i)]$
and we choose $g_{i';ij}$ to have the same spatial
dependence: $g_{i';ij}=g[\bm{r}_{i'}-(1/2)(\bm{r}_i+\bm{r}_j)]$.

Finally, the path integral representation of the
generating function for the dynamics is
$Z=1=\int \mathcal{D}[\phi,\bar\phi] \exp(-S[\phi,\bar\phi])$,
with
\begin{equation}
S[\phi,\bar\phi] = \int\! \mathrm{d}t\, \mathrm{d}^d\bm{r}\,
\bar{\phi}(\bm{r},t)
\partial_t \phi(\bm{r},t) + \int\! \mathrm{d}t\, H[\phi,\bar\phi,t] 
\label{equ:action}
\end{equation}
where 
\begin{eqnarray}
H[\phi,\bar\phi,t]&=& \int\!\mathrm{d}^d\bm{r}\,\mathrm{d}^d\bm{r}'\, 
g(\bm{r}-\bm{r'}) [1+\bar\phi(\bm{r'},t)] \phi(\bm{r'},t) \times \nonumber\\
& & \quad \Big\{ D
\nabla \bar{\phi}(\bm{r},t) \cdot \left[ (1+2\beta h(\bm{r},t)) 
\nabla \phi
(\bm{r},t) -
\phi(\bm{r},t) \beta \nabla h(\bm{r},t) \right]  \nonumber\\ & &
\quad\quad + \lambda 
\bar{\phi}(\bm{r},t) 
[2+\bar{\phi}(\bm{r},t)]
[1+2\beta h(\bm{r},t)] 
\phi(\bm{r},t)^2 \nonumber 
\\ & & \quad\quad + \mathcal{O}(\nabla^4,h^2) \Big\}
\label{equ:ham}
\end{eqnarray}
and $\lambda=\lambda_0 l_0^{2d}$, $D=D_0 l_0^{2+d}$

We work to linear order in the perturbation $h$ and to
quadratic order in a gradient expansion; we also
neglect boundary contributions to the action. 

This concludes our definition of the effective theory. We now
calculate the time dependence of the density, and the propagator
in the system.

\subsection{Mean density and propagator}

For the calculations of this subsection, we set the perturbing
field $h=0$. This field only enters the calculation of the
response, in the next subsection. 
To calculate the average density, we write
\begin{equation}
\partial_t n(t) \equiv \partial_t \langle \rho(\rbf,t) \rangle 
  = \langle \partial_t \phi(\rbf,t) \rangle.
\end{equation}
where the average is over trajectories with initial
density $n_0$, and weights given by the action of
(\ref{equ:action}).

Transforming to momentum space, we define
$\phi_{\bm{k}}(t)=\int\mathrm{d}^d\bm{r}\,
e^{i\bm{k}\cdot\bm{r}}\phi(\bm{r},t)$, and similarly 
$\bar\phi_{\bm{k}}(t)$. The Fourier transform of $g(\rbf)$
depends only on $k=|\bm{k}|$ and is denoted by $g_k$.
Working at tree level, 
we evaluate $\langle \partial_t \phi(\rbf,t) \rangle$
at the saddle point, whose position is given by
the Euler-Lagrange equation $\delta S / \delta{\bar{\phi}_{-\bm{q}}(t)}=0$.
Writing out this equation for $q=0$, we arrive at
\begin{equation}
\langle \partial_t \phi_{q=0}(t) \rangle 
= -\lambda \int\!\mathrm{d}^d\bm{r}\,\mathrm{d}^d\bm{r}'
 g(\bm{r}'-\bm{r}) 
 \Big\langle A[\bar\phi(t)]
\phi(\bm{r},t)^2 \phi(\bm{r}',t) \Big\rangle
\label{equ:eom_n}
\end{equation}
where $A[\bar\phi(t)]$ is a function that depends
on all of the Fourier components of $\bar\phi$, 
evaluated at a single time $t$.
To evaluate the right hand side of (\ref{equ:eom_n}),
we note that all expectation values of the form 
$\langle \bar\phi(t) B[\phi(t)] \rangle$ vanish (where
$B[\phi(t)]$ again denotes a function that is local
in time but depends on all Fourier components of 
the field $\phi$). 
Further, the tree level approximation is that 
$\langle B[\phi(t)] \rangle = B[\langle\phi(t)\rangle]$
for any function $B$. Hence, we arrive at
\begin{equation}
\partial_t n(t) = -2\lambda n(t)^3,
\end{equation}
where we used the fact that $g_{k=0}$ is 
normalised to unity, and $A[\bar\phi=0]=2$.
Starting from an initial condition with density $n_0$, we arrive
at the mean density:
\begin{equation}
n(t) = \frac{n_0}{\sqrt{1 + 4 \lambda n_0^2 t}} , 
\end{equation}

We now calculate the propagator for the dynamics, 
defined by
\begin{equation}
G_k(t,w)=(Nl_0^d)^{-1}
\langle \phi_{\bm{k}}(t) \bar\phi_{-\bm{k}}(w) \rangle,
\end{equation}
which satisfies $\lim_{w\to t^-} G_k(t,w)=1$. To calculate the
time dependence of $G_k(t,w)$, we 
obtain the time derivative
of $\bar\phi_{-\bm{k}}(t)$  from a
second Euler-Lagrange equation, 
$\delta S / \delta{\phi}_{\bm{k}}(t)=0$. 
At tree level, the only non-vanishing contributions 
to $\langle \phi_{\bm{k}}(t) \partial_w \bar\phi_{-\bm{k}}(w)\rangle$ 
are of the form
\begin{equation}
\langle \phi_{\bm{k}}(t) \bar\phi_{-\bm{k}}(w) U_k[\phi(w)] \rangle.
\end{equation}
(Terms of the form 
$\langle \phi_{\bm{k}}(t) \bar\phi_{-\bm{k}}(w) 
\bar\phi_{-\bm{k'}}(w) U'_k[\phi(w)] \rangle$ vanish since every
field $\bar\phi$ must be contracted with a field $\phi$ at a
later time.)
Making the only possible contraction, and evaluating
the function $U_k[\phi(w)]$ at the saddle point, we arrive at
\begin{equation}
\partial_w G_k(t,w)=G_k(t,w) U_k[\langle \phi(w) \rangle]
\end{equation}

Finally, we use the Euler-Lagrange equation to 
identify
$U_k[\langle \phi(w) \rangle]=Dk^2 n(w) + 2\lambda(2+g_k) n(w)^2$.
Hence:
\begin{equation}
G_k(t,w)=
\left( \frac{n(t)}{n(w)} \right)^{2+g_k} e^{z_k(w)-z_k(t)},
\end{equation}
with
\begin{equation}
z_k(t)=Dk^2/[2\lambda n(t)]
\label{equ:def_zk}
\end{equation}

Armed with expressions for $n(t)$ and $G_k(t,w)$ we are now
in a position to calculate the correlation and response.

\subsection{Correlation and response}

The two-point defect correlation function analogous to
$\tilde C_d(k,t,\tw)$ is
\begin{eqnarray}
C_\phi(k,t,\tw) &\equiv &
\int\mathrm{d}^d\bm{r}\, e^{i\bm{k}\cdot\bm{r}}
 [ \langle \rho(\bm{r},t) \rho(\bm{0},\tw)
\rangle - n(t) n(\tw) ].
\end{eqnarray}
The integrand decomposes as
\begin{equation}
e^{i\bm{k}\cdot\bm{r}}[ \langle \phi(\bm{r},t) \bar\phi(\bm{0},\tw) \phi(\bm{0},\tw)
 \rangle
+ \langle \delta\phi(\bm{r},t) \delta\phi(\bm{0},\tw) \rangle],
\label{equ:cphi_two}
\end{equation}
where $\delta\phi(\bm{r},t)=\phi(\bm{r},t)-n(t)$. This correlation
is to be evaluated at $h=0$.

At tree level, the integral of the
first term of (\ref{equ:cphi_two}) is simply 
$[G_k(t,\tw) n(\tw)]$. We can evaluate the second term by direct
consideration of its equations of motion. However, it is simpler
to identify the tree level contributions to this quantity directly:
expanding the time-ordered exponential $e^{-S}$, the only terms
contributing at tree-level come from:
\begin{equation}
\Big\langle \delta\phi(\bm{r},t) \delta\phi(\bm{r'},\tw) 
\int_0^{\tw}\!\mathrm{d}s \,
H[\phi,\bar\phi,s]
\Big\rangle.
\end{equation}
This expectation value can be calculated straightforwardly, by
making the allowed contractions
and using the tree level relation $\langle B[\phi(s)] \rangle
=B[\langle\phi(s)\rangle]$. The result is that
\begin{equation}
\langle \delta\phi_{-\bm{k}}(t) \delta\phi_{\bm{k}}(w) \rangle
=-2\lambda 
(Nl_0^d)
\int_0^w\mathrm{d}s\, G_k(t,s) G_k(w,s) n(s)^3 (1+2g_k)
\end{equation}
Making the change of variables, $z_s=Dk^2/[2\lambda n(s)]$,
leads to
\begin{eqnarray}
\langle \delta\phi_{-\bm{k}}(t) \delta\phi_{\bm{k}}(w) \rangle
&=&-
(Nl_0^d) 
n(w) G_k(t,w)
z_w^{-3-2g_k} \times \nonumber \\ & & \qquad
\int_{z_0}^{z_w}\! \mathrm{d}z_s\, 
 (1+2g_k) z_s^{2+2g_k} e^{2z_s-2z_w}
\end{eqnarray}
This tree level calculation is valid in the aging regime,
when $n(\tw)\ll n(t=0)$. In this limit, the integrand
is dominated by large $z_s$, and the behaviour is
independent of the initial condition. We therefore evaluate
the correlation at $z_0=0$.

The regimes of interest are $k\len\gg1$, for which $g_k\simeq0$,
and $k\len\ll 1$ for which $g_k\simeq1$. The integral can
be performed in these limits. Bringing everything together,
we arrive at
\begin{eqnarray}
C_\phi(k,t,\tw) = n(\tw) G_k(t,\tw)  F[z_k(\tw),g_k] 
\end{eqnarray}
where $z_k(t)$ was defined in (\ref{equ:def_zk}), and
\begin{eqnarray}
F[z,0]&=&1-(1/4)[2z^{-1}-2z^{-2}+z^{-3}(1-e^{-2z})] \\
F[z,1]&=&1-(3/4)[2z^{-1}-4z^{-2}+6z^{-3}-6z^{-4}+3z^{-5}(1-e^{-2z})] 
\end{eqnarray}
(These functions are regular at $z=0$,
which can be verified by direct expansion of the exponential.)

The calculation of the response is
also quite simple.
We define the impulse response by
\begin{eqnarray}
R_\phi(k,t,\tw) &=& \left. \frac{\delta}{\delta (\beta h_{\bm{k}}(\tw))} 
\langle \rho_{\bm{k}}(t) \rangle \right|_{h=0} \nonumber 
= - \Big\langle \phi_{\bm{k}}(t)
    \left.\frac{\delta S}{\delta (\beta h_{\bm{k}}(\tw))}\right|_{h=0}
    \Big\rangle
\end{eqnarray}
where $h_{\bm{k}}(\tw)$ is the Fourier transform of the
instantaneous perturbation $h(\bm{r})$ which acts
at time $\tw$. 

Substituting directly for the derivative of $S$, the 
terms that contribute at tree level are of the form
\begin{eqnarray}
R_\phi(k,t,\tw) &=& (Nl_0^d)^{-1} \langle \phi_{\bm{k}}(t) \bar\phi_{\bm{-k}}(\tw)
W[\phi(\tw)] \rangle,
\end{eqnarray}
where $\bar\phi_{\bm{-k}}(\tw)
W[\phi(\tw)]$ contains the terms in 
$\frac{\delta S}{\delta h_{\bm{k}}(\tw)}$
that are linear in $\bar\phi$.
Evaluating this
correlation function at the saddle point, 
we have 
$R_\phi(k,t,\tw) = G_k(t,\tw) W[\langle \phi(\tw)\rangle]$.
and hence
\begin{equation}
R_\phi(k,t,\tw) = 2\lambda[z_k(\tw)-2] n(\tw)^3 G_k(t,\tw)
\end{equation}

Evaluating the FDR by taking a derivative of the correlation,
we arrive at
\begin{eqnarray}
X_\phi(k,t,\tw) &\equiv& \frac{R_\phi(k,t,\tw)}{\partial_{\tw}C_\phi(k,t,\tw)}
\nonumber \\ &= & 
 \frac{-2+z_k(\tw)}{\left\{1+g_k+z_k(\tw)
[1+\frac{\mathrm{d}}{\mathrm{d}z_k(\tw)}]
\right\} F[z_k(\tw),g_k]}
\end{eqnarray}
where we note that the FDR is independent of $t$, and depends
on $k$ and $\tw$ only through the scaling variable $z_k(\tw)$.

As discussed in the main text,
the relevant limits for the SPM aging regime are $k\len\ll1$,
$k^2\gg \rho(\tw)$, and $\len^{-2} \ll k^2 \ll \rho(t)$. These
correspond to $[z_k(t)\ll1, g_k=0]$, $[z_k(\tw)\gg1, g_k=1]$, 
and $[z_k(t)\ll1,g_k=0]$ respectively. Calculating the FDR in these 
limits leads to
\begin{equation}
X_\phi(k,t,\tw) \equiv \frac{R_\phi(k,t,\tw)}{\partial_{\tw}C_\phi(k,t,\tw)}
= \left\{
\begin{array}{ccc}
  - \frac{5}{2} + \mathcal{O}(z_k(\tw)^{-1}) & & [k\len\ll1] \\
  1 + \mathcal{O}(z_k(\tw)) & & [k^2\gg \rho(\tw)] \\
  -3 + \mathcal{O}(z_k(\tw)^{-1}) & & [\len^{-2} \ll k^2 \ll \rho(t)] \\
\end{array}
\right.
\label{equ:X_field}
\end{equation}

As expected for a reaction-diffusion system with activated
dynamics \cite{Mayer06}, the large $k$ response has an
FDR of unity, while the response at small $k$ is negative. 
The intermediate regime with a well-defined
negative FDR is a result of the presence of two 
length scales in the SPM, both of which grow with time.


\section*{References}

\begin{thebibliography}{99}

\bibitem{ReviewsGT} For reviews see, e.g., M. D. Ediger, C. A. Angell
and S. R. Nagel, J. Phys. Chem. {\bf 100} 13200 (1996); C. A.  Angell,
Science {\bf 267}, 1924 (1995); P. G.  Debenedetti and F.H. Stillinger,
Nature {\bf 410}, 259 (2001).

\bibitem{Struik} L. C. E. Struik, {\em Physical Aging in Amorphous
Polymers and Other Materials}, (Elsevier, Houston, 1978).

\bibitem{Young} {\em Spin glasses \& random fields}, edited by
A.P. Young, (World Scientific, New York, 1998).

\bibitem{Garrahan} J. P. Garrahan, J. Phys. Condens. Matter {\bf 14},
1571 (2002).
 
\bibitem{BuhGar02} A. Buhot and J. P. Garrahan, Phys. Rev. Lett. {\bf
88}, 225702 (2002).

\bibitem{Jack05-spm} R. L. Jack, L. Berthier and J. P. Garrahan,
Phys. Rev. E {\bf 72}, 016103 (2005).

\bibitem{Jack-Garrahan} R. L. Jack and J. P. Garrahan,
J. Chem. Phys. {\bf 123}, 164508 (2005).

\bibitem{CugKur} L. F. Cugliandolo and J. Kurchan,
Phys. Rev. Lett. {\bf 71}, 173 (1993); J. Phys. A {\bf 27}, 5749
(1994).

\bibitem{laloux} J. Kurchan and L. Laloux, J. Phys. A {\bf 29}, 1929
  (1996).

\bibitem{CugKurPel97} L. F. Cugliandolo, J. Kurchan, and L. Peliti,
Phys. Rev. E {\bf 55}, 3898 (1997).

\bibitem{kurchan} J. Kurchan, Nature {\bf 433}, 222 (2005).

\bibitem{FraMezParPel98} S. Franz, M. M\'ezard, G. Parisi, and
L. Peliti, Phys. Rev. Lett. {\bf 81}, 1758 (1998).

\bibitem{CriRit03} A. Crisanti and F. Ritort, J. Phys. A {\bf 36},
R181 (2003).

\bibitem{FisHus86} D. S. Fisher and D. A. Huse, Phys. Rev. Lett. {\bf
56}, 1601 (1986).

\bibitem{ExpDH} See, for example: K. Schmidt--Rohr and H. Spiess,
Phys. Rev. Lett.  {\bf 66}, 3020 (1991); M. T. Cicerone and
M. D. Ediger, J. Chem. Phys. {\bf 103}, 5684 (1995); E. V. Russell and
N. E. Israeloff, Nature {\bf 408}, 695 (2000); E. Weeks et al., Science
{\bf 287}, 627 (2000); W. K. Kegel and A. van Blaaderen, Science {\bf
287}, 290 (2000); P. Mayer et al., Phys. Rev. Lett. {\bf 93}, 115701
(2004).

\bibitem{ReviewsDH} For reviews, see: H. Sillescu, J. Non-Cryst. Solids
                   {\bf 243}, 81 (1999); M. D. Ediger, Annu. Rev. Phys.
                   Chem. {\bf 51}, 99 (2000); S. C. Glotzer,
                   J. Non-Cryst.  Solids, {\bf 274}, 342 (2000);
                   R. Richert, J. Phys.  Condens. Matter {\bf 14},
                   R703 (2002); H. C. Andersen,
                   Proc. Natl. Acad. Sci. U. S. A. {\bf 102}, 6686
                   (2005).

\bibitem{Cri-Vio} A. Crisanti, F. Ritort, A. Rocco, and M. Sellitto,
J. Chem. Phys. {\bf 113}, 10615 (2000); P. Viot, J. Talbot, and
G. Tarjus, Fractals {\bf 11}, 185 (2003).

\bibitem{nicodemi} M. Nicodemi, Phys. Rev. Lett. {\bf 82}, 3734
(1999).

\bibitem{kr} F. Krzakala, Phys. Rev. Lett. {\bf 94}, 077204 (2005).

\bibitem{buhot} A. Buhot, J. Phys. A {\bf 36}, 12367 (2003).

\bibitem{Garrahan-Newman} J. P. Garrahan and M. E. J. Newman,
Phys. Rev. E {\bf 62}, 7670 (2000).

\bibitem{FieSol02} S. Fielding and P. Sollich, Phys. Rev. Lett. {\bf
88}, 050603 (2002).

\bibitem{MayBerGarSol} P. Mayer, L. Berthier, J. P. Garrahan, and
P. Sollich, Phys. Rev. E {\bf 68}, 016116 (2003); 
Phys. Rev. E {\bf 70}, 018102
(2004).

\bibitem{barrat} A. Barrat and L. Berthier, Phys. Rev. Lett. {\bf 87},
087204 (2001).

\bibitem{kennett} H. E. Castillo, C. Chamon, L. F. Cugliandolo, and
M. P. Kennett, Phys. Rev. Lett. {\bf 88}, 237201 (2002).

\bibitem{exp} L. Bellon, S. Ciliberto and C. Laroche, 
Europhys. Lett. {\bf 53}, 511 (2001). 

\bibitem{Mayer06} P. Mayer, S. Leonard, L. Berthier, J. P. Garrahan and
P. Sollich, Phys. Rev. Lett. {\bf 96}, 030602 (2006).

\bibitem{Lipowski} A. Lipowski, J. Phys. A {\bf 30}, 7365 (1997)

\bibitem{Newman-Moore} M. E. J. Newman and C. Moore, Phys. Rev. E {\bf
60}, 5068 (1999).

\bibitem{RitSol03} F. Ritort and P. Sollich, Adv. Phys. {\bf 52}, 219
(2003).

\bibitem{Other-plaquette} For studies of other glassy spin models with
plaquette interactions, which do not map simply to kinetically constrained
models, see for
example: D. Alvarez, S. Franz and F. Ritort, Phys. Rev. B {\bf 54},
9756 (1996); M. R. Swift, H. Bokil, R. D. M. Travasso and A. J. Bray,
Phys. Rev. B {\bf 62}, 11494 (2000); A. Lipowski and D. Johnston,
Phys. Rev. E {\bf 64}, 041605 (2001); P. Dimopoulos, D. Espriu,
E. Jane and A. Prats, Phys. Rev. E {\bf 66}, 056112 (2002).

\bibitem{CalGam} For a review, see P. Calabrese and A. Gambassi,
J.\ Phys. A {\bf 38}, R133 (2005).

\bibitem{EastModel} J. J\"ackle and S. Eisinger, Z. Phys. B {\bf 84},
  115 (1991).

\bibitem{nef1} L. Berthier and J.P. Garrahan,
J. Phys. Chem. B {\bf 109}, 3578 (2005).

\bibitem{nef2} D.J. Ashton, L.O. Hedges and J.P. Garrahan, 
J. Stat. Mech. P12010 (2005).

\bibitem{SolEvans} P. Sollich and M.R. Evans,
Phys. Rev. Lett. {\bf 83}, 3238 (1999); 
Phys. Rev. E {\bf 68}, 031504 (2003). 

\bibitem{FAModel} G.~H.~Fredrickson and H.~C.~Andersen,
Phys. Rev. Lett {\bf 53}, 1244 (1984).

\bibitem{Chatelain04} C.~Chatelain, J. Stat. Mech. P06006 (2004).

\bibitem{Ricci} F. Ricci-Tersenghi, Phys. Rev. E {\bf 68}, 065104
(2003).

\bibitem{coar1} A. Barrat, Phys. Rev. E {\bf 57}, 3629 (1998).

\bibitem{coar2} L. Berthier, J.-L. Barrat, and J. Kurchan,
Eur. Phys. J. B {\bf 11}, 635 (1999). 

\bibitem{FranzDPG99}
S. Franz, C. Donati, G. Parisi and S. C. Glotzer, Phil. Mag. B
{\bf 79}, 1927 (1999).

\bibitem{Mayer-prep} S. L\'eonard, P. Mayer, L. Berthier, 
J.P. Garrahan, and P. Sollich, in preparation.

\bibitem{SchTri99} M. Schulz and S. Trimper, J. Stat. Phys. {\bf 94},
173 (1999).

\bibitem{WhiBerGar04} S. Whitelam, L. Berthier, and J. P. Garrahan,
Phys. Rev. Lett. {\bf 92}, 185705 (2004); Phys. Rev. E {\bf 71},
026128 (2005); see also R. L. Jack, P. Mayer and P. Sollich
J. Stat. Mech (2006) P03006.

\bibitem{Lee94} B. P. Lee, J. Phys. A {\bf 27}, 2633
(1994).

\bibitem{tauber} U. C. T\"{a}uber, M. Howard,
and B. P. Vollmayr-Lee, J. Phys. A {\bf 38}, R79 (2005).

\bibitem{Doi-Peliti}
M. Doi, J. Phys. A {\bf 9}, 1479 (1976); L.~Peliti, J. Phys. (Paris)
{\bf 46}, 1469 (1984).

\end{thebibliography}
\end{document}